\documentclass[aps,prev,twocolumn,superscriptaddress,preprintnumbers,nofootinbib]{revtex4-1}
\pdfoutput=1

\usepackage{graphicx}
\usepackage{bm}
\usepackage{times}
\usepackage{slashed}
\usepackage{amssymb}
\usepackage{color}
\usepackage{aas_macros}
\usepackage{slashed}
\usepackage{lipsum}
\usepackage{subfigure}
\usepackage{multirow}
\usepackage{amsmath}
\usepackage{array}
\usepackage{varwidth}
\usepackage{float}

\newcommand{\be}{\begin{equation}}
\newcommand{\ee}{\end{equation}}
\newcommand{\bea}{\begin{eqnarray}}
\newcommand{\eea}{\end{eqnarray}}

\begin{document}


\title{Evidence for a New Component of High-Energy Solar Gamma-Ray Production}

\author{Tim Linden}
\email{linden.70@osu.edu; \; http://orcid.org/0000-0001-9888-0971}
\affiliation{Center for Cosmology and AstroParticle Physics (CCAPP), The Ohio State University, Columbus, OH 43210}

\author{Bei Zhou}
\email{zhou.1877@osu.edu; \; http://orcid.org/0000-0003-1600-8835}
\affiliation{Center for Cosmology and AstroParticle Physics (CCAPP), The Ohio State University, Columbus, OH 43210}
\affiliation{Department of Physics, The Ohio State University, Columbus, OH 43210}

\author{John F. Beacom}
\email{beacom.7@osu.edu; \; http://orcid.org/0000-0002-0005-2631}
\affiliation{Center for Cosmology and AstroParticle Physics (CCAPP), The Ohio State University, Columbus, OH 43210}
\affiliation{Department of Physics, The Ohio State University, Columbus, OH 43210}
\affiliation{Department of Astronomy, The Ohio State University, Columbus, OH 43210}

\author{Annika H. G. Peter}
\email{apeter@physics.osu.edu; \; http://orcid.org/0000-0002-8040-6785}
\affiliation{Center for Cosmology and AstroParticle Physics (CCAPP), The Ohio State University, Columbus, OH 43210}
\affiliation{Department of Physics, The Ohio State University, Columbus, OH 43210}
\affiliation{Department of Astronomy, The Ohio State University, Columbus, OH 43210}

\author{Kenny C. Y. Ng}
\email{chun-yu.ng@weizmann.ac.il; \; http://orcid.org/0000-0001-8016-2170}
\affiliation{Department of Particle Physics and Astrophysics, Weizmann Institute of Science, Rehovot 76100, Israel}

\author{Qing-Wen Tang}
\email{qwtang@ncu.edu.cn; \; http://orcid.org/0000-0001-7471-8451}
\affiliation{Center for Cosmology and AstroParticle Physics (CCAPP), The Ohio State University, Columbus, OH 43210}
\affiliation{Department of Physics, Nanchang University, Nanchang 330031, China}

\begin{abstract}
The observed multi-GeV gamma-ray emission from the solar disk --- sourced by hadronic cosmic rays interacting with gas, and affected by complex magnetic fields --- is not understood. Utilizing an improved analysis of the Fermi-LAT data that includes the first resolved imaging of the disk, we find strong evidence that this emission is produced by two separate mechanisms. Between 2010--2017 (the rise to and fall from solar maximum), the gamma-ray emission is dominated by a polar component. Between 2008--2009 (solar minimum) this component remains present, but the total emission is instead dominated by a new equatorial component with a brighter flux and harder spectrum. Most strikingly, although 6 gamma rays above 100 GeV are observed during the 1.4 years of solar minimum, none are observed during the next 7.8 years.  These features, along with a 30--50 GeV spectral dip which will be discussed in a companion paper, were not anticipated by theory.  To understand the underlying physics, Fermi and HAWC observations of the imminent Cycle 25 solar minimum are crucial.  
\end{abstract}

\maketitle

The Sun is a bright source of multi-GeV $\gamma$-rays, with emission observed both from its halo --- due to cosmic-rays electrons interacting with solar photons --- and its disk --- due to hadronic cosmic rays (mostly protons) interacting with solar gas.  
(Emission from solar particle acceleration is only bright during flares and has not been observed above 4 GeV~\cite{1987ApJS...63..721M, 1988SSRv...49...69K, 1993A&AS...97..349K, Ackermann:2013tya, Pesce-Rollins:2015hpa, Ackermann:2017uer, Share:2017tgw, Fermi-LAT:2013cla}.)  Although the halo emission~\cite{2011ApJ...734..116A} agrees with theory~\cite{2006ApJ...652L..65M, 2007Ap&SS.309..359O, Orlando:2017iyc}, the disk emission does not, and hence is our focus.

Until recently, the most extensive analysis of solar disk $\gamma$-ray emission was based on Fermi-LAT data from 2008--2014~\cite{Ng:2015gya} (for earlier work, see Refs.~\cite{Orlando:2008uk, 2011ApJ...734..116A}), and produced three results. First, the flux is bright, e.g., at 10 GeV, it exceeds the flux expected from Earth-directed cosmic rays interacting with the solar limb by a factor $\gtrsim$50~\citep{Zhou:2016ljf}. Second, it continues to 100 GeV, requiring proton energies $\sim$1000~GeV.  Third, the 1--10 GeV flux is anti-correlated with solar activity, and is $\sim$2.5$\times$ larger at solar minimum than maximum.  The {\it only} theoretical model of disk emission is the 1991 paper of Seckel, Stanev, and Gaisser (SSG)~\cite{Seckel:1991ffa}, which proposes that magnetic flux tubes can reverse incoming protons deep within the solar atmosphere, where they have an appreciable probability of producing outgoing $\gamma$-rays.  Even though this enhances the $\gamma$-ray flux, the SSG prediction still falls a factor $\sim$6 below the data at 10 GeV, and does not explain the time variation. 

Now, in this and an upcoming companion paper, we perform new analyses of Fermi-LAT data based on a longer exposure (now 2008--2017), better data quality (Pass 8), and improved methods.  In Ref.~\cite{paper}, we focus on the 1--100 GeV spectrum and its time variability.  Compared to our earlier work in Ref.~\cite{Ng:2015gya}, the flux is detected more robustly up to 100 GeV, and the anticorrelation of the flux with solar activity is detected up to $\sim$30~GeV.  Most significantly, we discover a spectral dip between 30--50 GeV.  This dip is unexpected and its origin is unknown.  Here we extend the analyses of Refs.~\cite{Ng:2015gya, paper} by going to higher energies, studying the time variation in a new way, and performing the first analysis of flux variations across the resolved solar disk.  In the following, we detail our methodology, highlight key discoveries, and discuss their possible theoretical implications.

The importance of this work is manifold.  Because the disk $\gamma$-ray emission is brighter and more mysterious than expected, it motivates new searches with Fermi~\citep{Atwood:2009ez}, the higher-energy HAWC $\gamma$-ray experiment~\citep{Abeysekara:2013tza}, and the IceCube neutrino observatory~\citep{Aartsen:2012kia}.  The results will yield valuable insights on the complex, dynamic solar magnetic environment, from cosmic-ray modulation in the solar system to the fields deep within the photosphere.  They will also advance searches for new physics~\cite{Batell:2009zp, Schuster:2009au, Meade:2009mu, Bell:2011sn, Feng:2016ijc, Leane:2017vag, Arguelles:2017eao, Edsjo:2017kjk, Ng:2017aur}.  Most generally, these searches provide the highest-energy data available in the program to understand the Sun as an example of other stars. \newline

\noindent \emph{Methodology.---} We utilize front and back Pass 8 Source events from August 4, 2008 to November 5, 2017 (MET: 239557417--531557417), employing standard cuts. We include events exceeding 10~GeV observed within 0.5$^\circ$ of the solar center (the Sun's angular radius is 0.26$^\circ$). The excellent angular resolution of $>$10~GeV $\gamma$-rays minimizes the flux lost from our ROI. In Appendix~\ref{appendix:angularcut}, we show that using a larger ROI does not affect our results. We remove events observed while the Sun falls within 5$^\circ$ of the Galactic plane, due to the larger diffuse background. This cut is smaller than in previous work, but is sufficient due to the small ROI. We perform the first conversion of each $\gamma$-ray to Helioprojective coordinates utilizing {\tt sunpy}~\citep{2015CS&D....8a4009S} and {\tt astropy}~\citep{2013A&A...558A..33A}. We ignore diffuse backgrounds, which we found in Ref.~\citep{paper} to be negligible.

We calculate the Fermi-LAT exposure at the solar position in temporal bins of 5000~s (but use precise photon times for recorded events).  Within this period, the Sun moves less than 0.1$^\circ$ in the Fermi coordinate system, and the Fermi-LAT effective exposure is approximately constant. We assume a single effective exposure over the full ROI in each time-bin, and bin the exposure into 32 logarithmic energy bins spanning 10~GeV to 1~TeV. Because the Sun occupies a unique position in instrumental $\phi$-space, we calculate exposures obtained by utilizing 10 independent $\phi$-bins. In Appendix~\ref{appendix:phidependence}, we show that the instrumental $\phi$-dependence does not affect our results. \newline

\begin{figure}[tbp]
\centering
\includegraphics[width=0.48\textwidth]{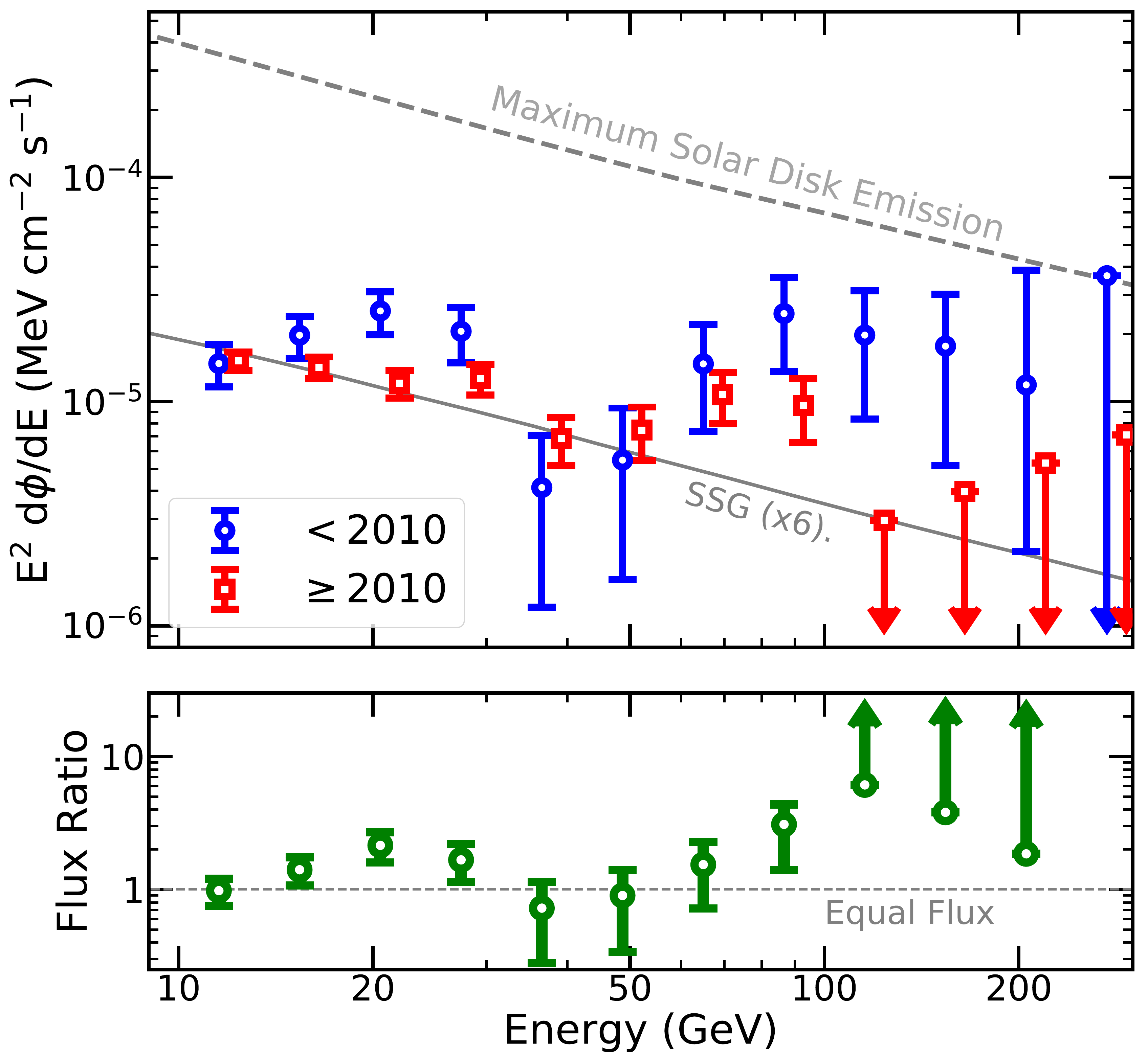}
\caption{ {\bf (Top)} The solar disk $\gamma$-ray spectrum during solar minimum (before January 1, 2010, blue circles) and after (red squares). Small shifts along the x-axis improve readability. The gray lines show the SSG model renormalized by a factor of six to fit the lowest-energy datapoint (solid), and the maximum $\gamma$-ray flux that could be produced by hadronic cosmic rays (dashed). {\bf (Bottom)} The ratio of the $\gamma$-ray flux observed in periods during and after solar minimum.}
\label{fig:spectrumratio}
\end{figure}

\noindent \emph{Flux, Spectrum and Time Variation.---}In Figure~\ref{fig:spectrumratio}, we show the solar $\gamma$-ray flux before and after January 1, 2010, which roughly corresponds to the end of the Cycle 24 solar minimum. We note three key results.

\begin{itemize}
\item  The $\gamma$-ray flux significantly exceeds the SSG prediction (based on a proton interaction probability of $0.5\%$), in fact approaching the maximum allowed solar disk flux (for a detailed calculation, see Appendix~\ref{appendix:maximum}).

\item The 30--50~GeV spectral dip, which we will carefully examine in Ref.~\citep{paper}, is statistically significant both during and after solar minimum, though there is some evidence (2.5$\sigma$) that the dip deepens at solar minimum. Aside from the dip, the spectra in both time periods are significantly harder than predicted by SSG.

\item The strongest time variation is observed between solar minimum (largest flux), and the remaining solar cycle. At low energies this variation is moderate~\citep{Orlando:2008uk, Ng:2015gya, paper}. However, the amplitude increases with energy above 50~GeV, reaching a factor $\geq$10 above 100~GeV.
\end{itemize}
\noindent None of these observations were anticipated by theory.

\noindent \emph{Morphology.---}The large $\gamma$-ray flux suggests that a large fraction of the solar surface participates in the $\gamma$-ray emission process. To further elucidate the $\gamma$-ray generation mechanism(s), we resolve the $\gamma$-ray morphology across the solar surface. This reconstruction is possible at high ($\gtrsim$10~GeV) energies due to the excellent ($\sim$0.1$^\circ$) Fermi angular resolution. 

\begin{figure*}[!htp]
\centering
\includegraphics[width=0.9\textwidth]{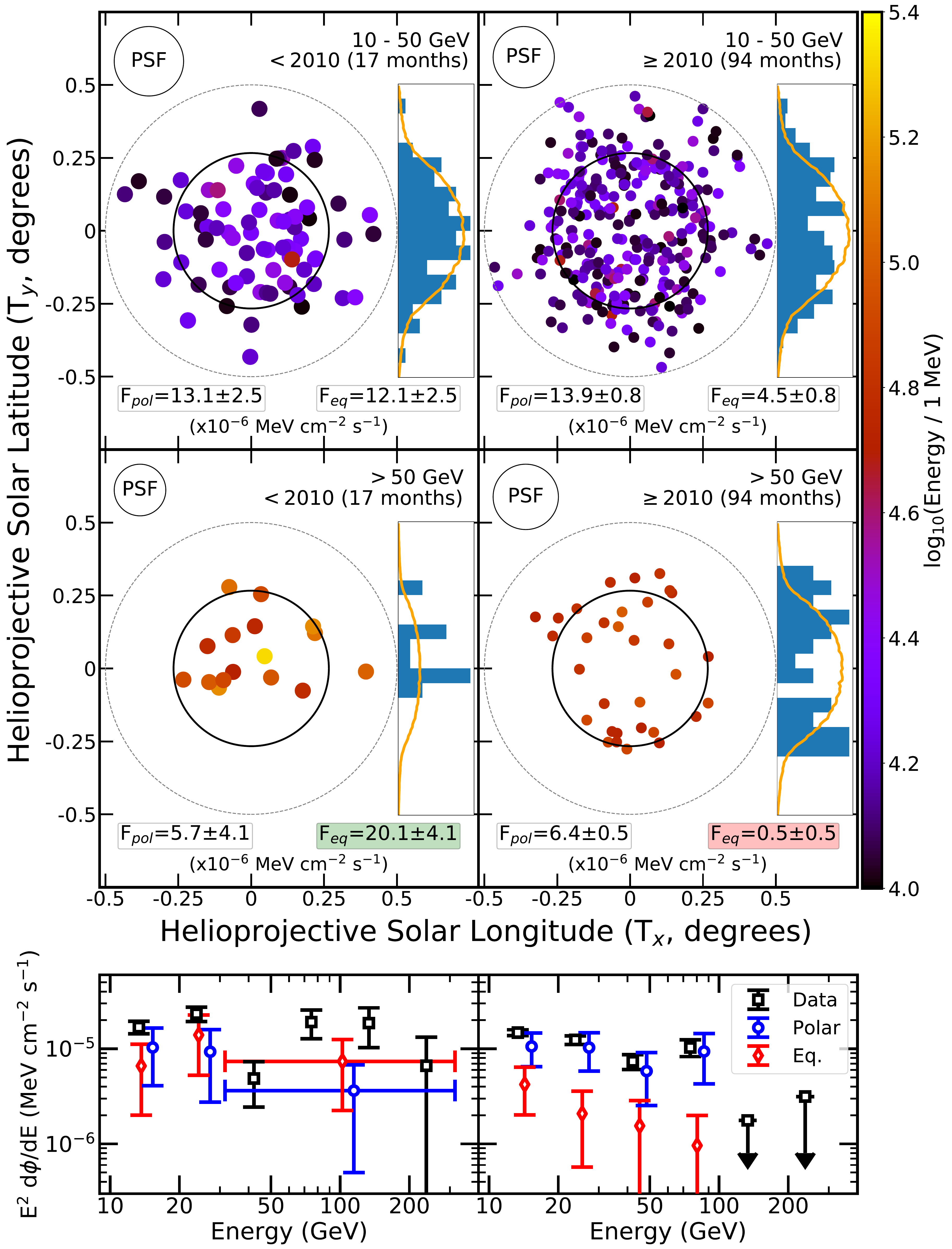}
\caption{ {\bf (Top)} The location and energy of solar $\gamma$-rays in Helioprojective coordinates. Data are cut into two temporal bins and two energy bins. The solar disk is represented by the solid circle, and the 0.5$^\circ$ ROI by the dashed circle. The average PSF of observed $\gamma$-rays is depicted in the top left. The T$_y$ positions of photons are shown in the histogram, and are compared to the profile expected from isotropic emission smeared by the PSF (orange line). The area of event points corresponds to the relative effective area in data taken during (after) solar minimum. In each bin, we report the flux from the modeled polar and equatorial components, as described in the text. {\bf (Bottom)} The energy spectrum of polar and equatorial emission, divided into regions during (left) and after (right) solar minimum. The polar emission is approximately constant, while the equatorial emission decreases drastically at the end of solar minimum.}
\label{fig:helioprojective}
\end{figure*}

In Figure~\ref{fig:helioprojective}, we show the observed position of $\gamma$-rays in our analysis, dividing the data into two temporal bins (before and after January 1, 2010; corresponding to the end of the solar minimum), and two energy bins (below and above 50~GeV; corresponding to the spectral dip discussed in Ref.~\citep{paper}).  Surprisingly, we find that, contrary to the SSG model, the emission is neither isotropic nor time-invariant. Instead, it includes distinct polar and equatorial components, with separate time and energy dependences. In particular, it is visually apparent that $\gamma$-rays above 50~GeV are predominantly emitted near the solar equatorial plane during solar minimum, but are emitted from polar regions during the remaining solar cycle.

We utilize two separate methods to quantify the significance of this morphological shift. The first employs a Kolmogorov-Smirnov test to differentiate the distribution of $\gamma$-rays in observed helioprojective latitude ($|T_y|$) during and after solar minimum. This provides a model independent method of comparing the data, but loses sensitivity to convolving factors such as the instrumental PSF. Below 50~GeV, we find that the event morphology is consistent to within 1.1$\sigma$. However, above 50~GeV, we reject the hypothesis that the event morphologies during and after solar minimum are equivalent at 2.8$\sigma$. Because this method has few trials, it provides reasonable evidence for a morphological shift.

Second, we define a two-component model of the solar surface, with equal-area equatorial and polar emission components (divided at T$_y$~=~$\pm$0.108$^\circ$). We fit the flux from each component, utilizing the angular reconstruction of each observed $\gamma$-ray (see Appendix~\ref{appendix:morphologymodeling}). This correctly accounts for the PSF, but provides results that depend on the assumed emission model. In Appendix~\ref{appendix:limbbackground} we show that different models produce similar results. This analysis provides two key results.

\begin{itemize}
\item At all energies, the $\gamma$-ray emission becomes more polar after solar minimum. However, the amplitude of this shift increases significantly at high energies.

\item The morphological shift is produced by a significant decrease in the equatorial flux after solar minimum, while the polar flux remains relatively constant.
\end{itemize}

In Figure~\ref{fig:helioprojective}, we also plot the polar and equatorial spectra during and after solar minimum. We find that while the amplitude and spectrum of the polar component remains relatively constant, the equatorial spectrum softens substantially after solar minimum. This significantly decreases the high-energy equatorial flux after solar minimum, despite the similar normalization of the equatorial component at low energies. Intriguingly, the equatorial $\gamma$-ray spectrum during solar minimum is extremely hard, and is consistent with dN/dE$\sim$E$^{-2}$ up to energies significantly exceeding 100~GeV. We note that we have combined high-energy spectral bins during solar minimum to provide sufficient statistics.

\begin{table*}[!t]
\begin{tabular}{| c | c | c | c | c | c | c | c | c | c | c |}
\hline\hline
Time (UTC) & Energy & R.A. & Dec & Solar Distance & Event Class & PSF Class & Edisp Class & P6 & P7 & BG Contribution \\ \hline
2008-11-09 03:47:51 & 212.8 GeV & 224.497 & -16.851 & 0.068$^\circ$ & UltraCleanVeto & PSF0 & EDISP3 & \checkmark & \checkmark & 0.00050 \\
\hline
2008-12-13 03:25:55 & 139.3 GeV & 260.707 & -23.243 & 0.126$^\circ$ & UltraCleanVeto & PSF2 & EDISP1 & X & X & 0.00038 \\
\hline
2008-12-13 07:04:07 & 103.3 GeV & 260.346 & -23.102 & 0.399$^\circ$ & UltraCleanVeto & PSF0 & EDISP2 & X & X & 0.00052 \\
\hline
2009-03-22 08:43:13 & 117.2 GeV & 1.337 & 0.703 & 0.255$^\circ$ & UltraCleanVeto & PSF1 & EDISP3 & \checkmark & \checkmark & 0.00027\\
\hline
2009-08-15 01:14:17 & 138.5 GeV & 144.416 & 14.300 & 0.261$^\circ$ & UltraCleanVeto & PSF2 & EDISP3 & \checkmark & \checkmark & 0.00021 \\
\hline
2009-11-20 07:55:20 & 112.6 GeV & 235.905 & -19.473 & 0.288$^\circ$ & UltraCleanVeto & PSF1 & EDISP1 & X & X & 0.00020 \\
\hline
\hline
\hline
2008-12-24 05:41:53 & 226.9 GeV & 272.899 & -23.343 & 0.069$^\circ$ & UltraClean & PSF1 & EDISP3 & X & X & 0.00128\\
\hline
2009-12-20 08:06:31 & 467.7 GeV & 268.046 & -23.177 & 0.338$^\circ$& UltraCleanVeto & PSF1 & EDISP0 & X & X & 0.00208 \\
\hline
\hline
\end{tabular}
\caption{Event Information for P8R2\_SOURCE\_V2 events with recorded energies exceeding 100~GeV observed within 0.5$^\circ$ of the solar center. Checkmarks indicate events that were recorded as photons in previous Pass 6 and Pass 7 analyses, while the Background contribution indicates the probability that diffuse emission produced the event. Events below the double-line did not pass our default selection criteria, as they were observed when the Sun was located within 5$^\circ$ of the Galactic plane.} 
\label{tab:100GeVphotons}
\end{table*}

\noindent \emph{Flux Above 100~GeV.---} In Figure~\ref{fig:spectrumratio}, we discovered a bright $\gamma$-ray flux above 100~GeV during solar minimum, but found no events in the remaining solar cycle. In Table~\ref{tab:100GeVphotons}, we provide detailed information concerning each $>$100~GeV event in our analysis. We uncover no significant concerns regarding the event classes, or angular and energy reconstructions. In particular, all six events pass the UltraCleanVeto event cut, providing the highest confidence that they are true $\gamma$-rays. We calculate the probability that each event has a non-solar origin by calculating the $\gamma$-ray flux above 100~GeV in each ROI during periods when Sun is not present. We find that diffuse contributions cannot explain these events. The total diffuse $\gamma$-ray flux above 100~GeV over the solar path produces $\sim$0.3 background event over the full analysis period (see Appendix~\ref{appendix:fakesuns}). 

Examining each event yields three insights. First, we observe several extremely high-energy events, including three events exceeding 200~GeV, and one event at 470~GeV. This suggests that multi-TeV protons can produce outgoing $\gamma$-rays through solar interactions, and that HAWC observations of the upcoming solar minimum may be illuminating. 

Second, all six events in our default analysis were observed between November 2008 and November 2009, which is inconsistent with a steady-state hypothesis. We determine the significance of this temporal variability by conducting a Kolmogorov-Smirnov test of the hypothesis that the data is Poissonian in solar exposure. We rule out the steady-state hypothesis at 4.2$\sigma$. Noting that the Sun moves through the Galactic plane during solar minimum and that the diffuse contribution along the plane remains negligible above 100~GeV, we unblind the region b~$<$~5$^\circ$. This adds two new events above 100~GeV that were also observed at solar minimum and no additional events during the subsequent eight passages of the Galactic plane. Including these events increases the statistical significance for temporal variability to 4.9$\sigma$.

Third, we note a peculiar ``double-event" occurring on December 13, 2008, when two $>$100~GeV $\gamma$-rays were observed within 3.5~hr. The probability that any two events are this closely correlated is inconsistent with the Poissonian expectation at $\sim$2.9$\sigma$. Intriguingly, the double event occurred during a significant solar-minimum coronal mass ejection event which began on December 12, 2008 and encountered Earth on December 17, 2008~\citep{Byrne:2010rz, Byrne:2012nk, 2013ApJ...769...43D}.

\noindent \emph{Interpretation.---}We have shown several lines of evidence that reveal two distinct high-energy $\gamma$-ray emission components on the solar disk. The first emits primarily from the Sun's polar regions, has a constant amplitude over the solar cycle, and produces no observed flux above 100~GeV. The second emits primarily from the Sun's equatorial plane, has an amplitude that decreases drastically after solar minimum, and has a hard spectrum at solar minimum that extends above 200~GeV.

These results are not explained by the SSG model. The bright $\gamma$-rays flux across the solar surface does support the SSG mechanism of cosmic-ray reversal deep within the photosphere. However, the flux, spectrum, time-variation, morphological shift, and spectral dip of solar $\gamma$-rays are unexplained. We can qualitative parameterize the solar $\gamma$-ray flux as:  

\vspace{-0.3cm}
\begin{equation}
\label{eq:1}
\Phi_{\rm \odot}(E_{\rm \gamma}) = \pi R_\odot^2 \Phi_{\rm CR}(E_{\rm CR}) C(E_{\rm \gamma}, E_{\rm CR}) f_{{\rm sur}}f_{{\rm turn}}f_{\rm int}
\end{equation} 

\vspace{-0.1cm}
\noindent where $\Phi_\odot$ is the disk $\gamma$-ray flux, $\Phi_{CR}$ is the cosmic-ray flux at the solar surface, $C$ describes the $\gamma$-ray flux at energy E$_{\rm \gamma}$ produced by a hadronic interaction at energy E$_{\rm CR}$ (see Appendix~\ref{appendix:maximum}), f$_{{\rm sur}}$ is the fraction of the solar surface that produces $\gamma$-rays, f$_{{\rm turn}}$ is the fraction of incoming cosmic rays that are reversed by magnetic fields within the solar photosphere, and f$_{\rm int}$ is the fraction of these cosmic rays that undergo a hadronic interaction and produce outgoing $\gamma$-rays before leaving the surface. SSG found solar modulation to be a small effect, implying that $\Phi_{\rm CR}$ is similar to the interstellar cosmic-ray flux. SSG assumes that each efficiency term is energy, position, and time independent. In particular, SSG set f$_{{\rm sur}}$ and f$_{{\rm turn}}$ to unity, and calculated f$_{{\rm int}}\sim$~0.5\%. 

Our observations instead indicate that these parameters strongly depend on the cosmic-ray energy, solar cycle, and solar latitude. Focusing on solar minimum, these shifts are more remarkable for four reasons. The large flux, within a factor of $\sim$4 of the maximal value, implies that all efficiency parameters are near unity. The hard spectrum, significantly exceeding the E$^{-2.7}$ interstellar cosmic-ray spectrum, indicates that these efficiencies rise quickly with energy. The equatorial morphology indicates that polar regions are not emitting efficiently, implying $f_{\rm sur}$~$\lesssim$0.5. Finally, symmetry constrains $f_{\rm int}$~$\sim$~0.5, as cosmic rays should undergo equal interactions while entering and exiting the photosphere. These observations produce significant tension with any SSG-like model.

This tension motivates us to consider scenarios that violate the assumptions of Eq.~(\ref{eq:1}) and allow for larger $\gamma$-ray fluxes.  First, cosmic rays may be collected from a larger area than the physical disk.  This is suggested by Tibet AS$\gamma$ observations of the Sun's cosmic-ray shadow~\citep{Amenomori:2013own}, which found that the area of the 10~TeV shadow was enhanced by $\sim$50\% during the Cycle-23 minimum. These effects may be larger at low energies. However, this would further soften the (already too soft) predicted $\gamma$-ray spectrum. Second, cosmic-ray capture may not be in equilibrium with $\gamma$-ray production, producing periods of enhanced $\gamma$-ray emission. However, we find no delay between solar minimum and the enhanced $\gamma$-ray flux.  Third, cosmic rays secondaries that exit the photosphere in the initial interaction may re-encounter the Sun. However, the total $\gamma$-ray production from multiple primary transits is taken into account by SSG, and the extra energy in secondary cosmic rays is only $\sim$35\%. Finally, cosmic-rays may be collected from all angles, while $\gamma$-rays are preferentially emitted perpendicular to the solar rotation axis. However, the magnetic field geometry capable of producing this anisotropy is unknown. 

One potential insight stems from the two correlated $>$100~GeV $\gamma$-rays observed on December 13, 2008. These events may be connected to a contemporaneous Earth-bound coronal mass ejection (CME) that began on December 12, 2008 and encountered Earth on December 17~\citep{Byrne:2010rz, Byrne:2012nk, 2013ApJ...769...43D}. When the $\gamma$-rays were observed, the CME had propagated to $\sim$40~R$_\odot$, implying that the $\gamma$-ray emission is not produced through collisions with the low-density ejecta. However, the significant magnetic field enhancements and open field lines associated with CMEs~\citep{1985JGR....90..275I, 2011JASTP..73.1129P} provide locations capable of reversing TeV cosmic rays. Additionally, CMEs occur predominantly along the solar equatorial plane during solar minimum, while they are more isotropic at solar maximum~\citep{2003ApJ...586..562G, 2004JGRA..109.7105Y} potentially explaining the time variation in the disk $\gamma$-ray morphology. However, none of the remaining $>$100~GeV $\gamma$-rays correspond to significant CMEs. Moreover, the covering fraction of CMEs on the solar disk is small, while the bright solar disk flux implies that any mechanism must be active across a large fraction of the solar surface.

Two other phenomena may explain the morphological shifts in our data. The solar $\gamma$-ray flux may be enhanced through interactions in helmet streamers, which are the largest closed loops present near active solar regions~\citep{1996SoPh..167..217L}. Helmet streamers have large magnetic fields and high gas densities that would be capable of trapping and converting high-energy cosmic-rays. Like CMEs, helmet streamers are observed along the solar equatorial plane at solar minimum, but are homogeneous during the remaining solar cycle~\citep{2004ApJ...603..307E}. On the other hand, the $\gamma$-ray flux may be inhibited in regions with coronal holes, which are the open field lines connecting the photosphere to the interplanetary magnetic field. The strong, ordered magnetic fields in coronal holes may prevent cosmic-rays from reaching deep into the photosphere. During solar minimum, coronal holes are found in polar regions, while they are primarily equatorial at solar maximum~\citep{2016ApJ...827L..41F}. However, we note that neither of these mechanism can produce a $\gamma$-ray flux exceeding the unity efficiency assumed in our maximal model, thus they are likely to be only part of the story.

\noindent \emph{A New Event!---}While finalizing this letter, we found a new $>$100~GeV event. Observed on February 13, 2018 at 17:49:15 UTC, the event has an energy of 162~GeV, is located 0.36$^\circ$ from the solar center, passes the UltraCleanVeto event selection, and belongs to the PSF0 and EDISP3 event classes. As we re-enter solar minimum, this is the first $>$100~GeV event recorded within 0.5$^\circ$ of the sun since 2009. The event may be connected to a Earth-bound CME observed on February 12, 2018.\footnote{https://www.swpc.noaa.gov/news/c1-flare-and-associated-cme-12-february-2018} Preliminary work indicates that this event increases the significance of the $>$100~GeV time variability above 5$\sigma$, and provides evidence that the upcoming solar minimum will provide a substantial flux of high-energy events. \newline

\noindent \emph{Future Outlook.---}We have discovered statistically significant temporal variations in the intensity, spectrum and morphology of solar $\gamma$-ray emission. These variations strongly suggest that two distinct components substantially contribute to the total solar $\gamma$-ray flux, including (1) a polar component that varies moderately in time and has a $\gamma$-ray spectrum that falls sharply around 100~GeV, and (2) an equatorial component with an extremely hard $\gamma$-ray spectrum that continues above 200~GeV, but is dominant only during solar minimum. These observations provide important new clues about the mechanisms behind solar disk $\gamma$-ray emission, which remains mysterious.

This mystery is deepened by the high intensity and hard spectrum of disk emission. In particular, the solar minimum flux appears to be in tension with the most optimistic predictions from the class of models that convert the interstellar cosmic-ray flux into a time-invariant and isotropic $\gamma$-ray flux. If future observations detect emission at even moderately higher energies, a new theoretical mechanism will be necessary to explain the highest-energy solar emission. 

Fortunately, observations of the upcoming Cycle 25 solar minimum by both the Fermi-LAT and HAWC will provide valuable information. Preliminary estimates indicate that the Cycle 25 minimum will be even quieter than the already quiet Cycle 24 minimum~\citep{2016JGRA..12110744H}. The observation of $>$100~GeV $\gamma$-rays during this period will significantly enhance our understanding of high-energy $\gamma$-ray emission from the solar disk. One new event has recently been detected. With improved statistics, it will soon become possible to correlate $\gamma$-ray events with solar observables, shining light on the magnetic field processes responsible for the high-energy $\gamma$-ray flux.

\vspace{-0.6cm}
\section*{A\MakeLowercase{cknowledgements}}
\vspace{-0.5cm}
We thank Joe Giacalone, Dan Hooper, Igor Moskalenko, Nick Rodd, Andy Strong, and especially Keith Bechtol, Ofer Cohen, and Stuart Mumford for helpful comments. This research makes use of SunPy, an open-source, community-developed solar analysis package~\citep{2015CS&D....8a4009S}. TL, BZ, and AHGP are supported in part by NASA Grant No. 80NSSC17K0754. BZ is also supported by a University Fellowship from The Ohio State University. JFB and BZ (partially) are supported by NSF grant PHY-1714479. KCYN is supported by the Croucher Fellowship and Benoziyo Fellowship.

\newpage
\appendix

\section*{Supplemental Material}

Here, we discuss numerous additional tests of our data analysis techniques and selection cuts. We first focus on systematic issues that could affect the determination of either the $>$100~GeV solar $\gamma$-ray flux or the solar $\gamma$-ray morphology (Appendices~\ref{appendix:angularcut}--\ref{appendix:pass6pass7}). We then present an improved calculation of the maximum solar $\gamma$-ray emission expected from an \mbox{SSG-like} model (Appendix~\ref{appendix:maximum}).  Next, we focus on alternative cuts of the $\gamma$-ray data that motivate the temporal and angular cuts utilized throughout the paper (Appendices~\ref{appendix:morphologymodeling}--\ref{appendix:flip}). Finally, we examine whether the 30-50~GeV spectral dip is confined only to events observed from polar or equatorial regions of the solar surface, but find that the photon statistics are insufficient to make any such statement (Appendix~\ref{appendix:polardip}). 

\section{ROI Dependence of the Solar $\gamma$-ray Spectrum}
\label{appendix:angularcut}

In our main analysis, we use a 0.5$^\circ$ region of interest surrounding the solar center. This cut improves our analysis in two ways. First, it eliminates most of the $\gamma$-ray background, because the $>$10~GeV solar disk flux of $\sim$2$\times$10$^{-5}$~MeV~cm$^{-2}$~s$^{-1}$ significantly exceeds all but the brightest sources in the $\gamma$-ray sky, and additionally because the Sun is physically blocking 25\% of the diffuse $\gamma$-ray emission within the ROI. Second, the 0.5$^\circ$ angular cut significantly decreases the flux from the inverse-Compton scattering of solar photons to $\gamma$-ray energies. While this component has a similar intensity as solar disk $\gamma$-rays, it is extended over $\sim$10$^\circ$. Furthermore, the inverse-Compton component is kinematically suppressed by Klein-Nishina effects across the solar disk. Thus, the choice of a small ROI renders the inverse-Compton contribution negligible. 

\begin{figure}[tbp]
\centering
\includegraphics[width=0.48\textwidth]{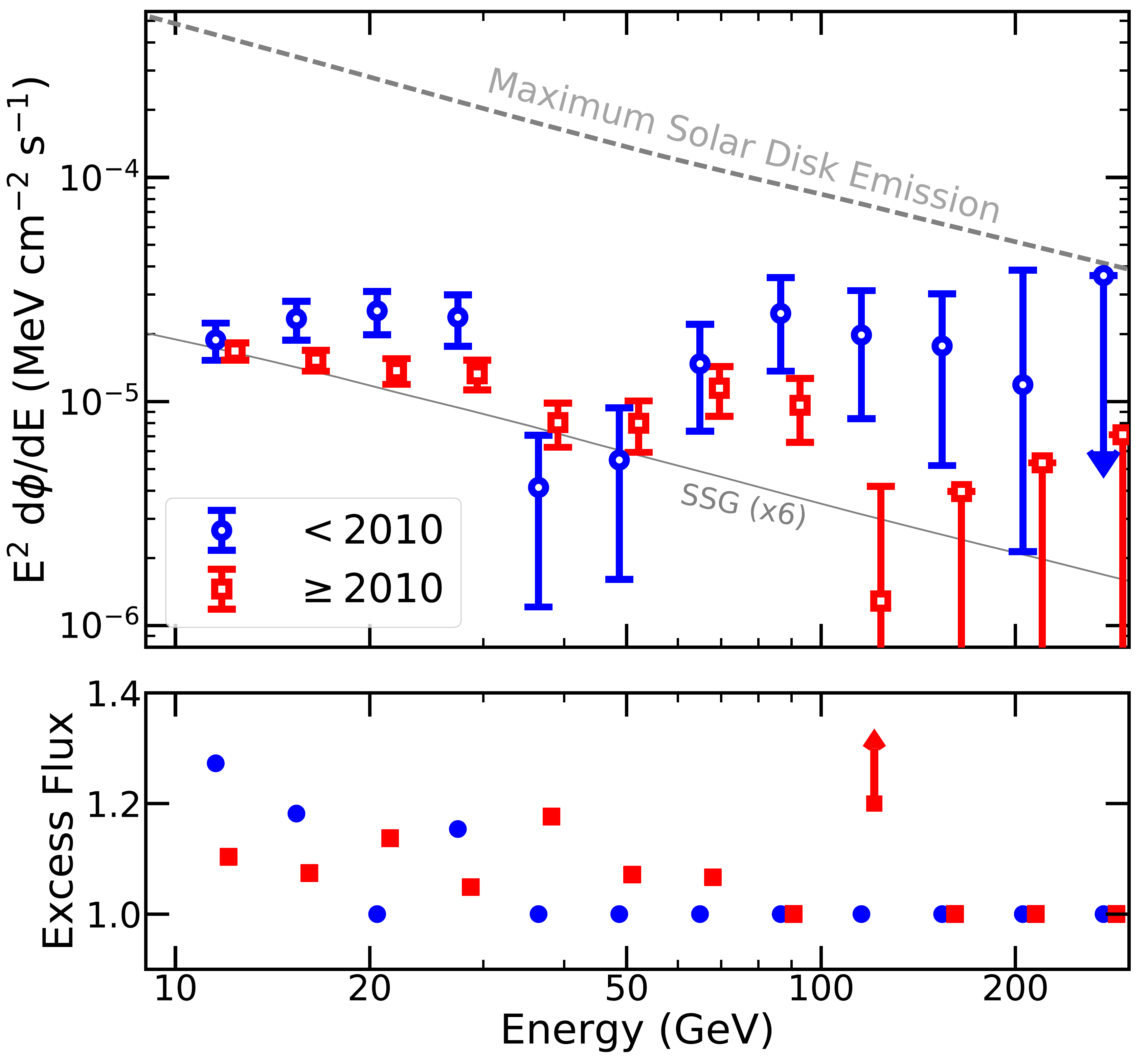}
\caption{ {\bf (Top)} Same as Figure~\ref{fig:spectrumratio}, but utilizing a larger angular cut of 0.75$^\circ$ on the solar  $\gamma$-ray flux. {\bf (Bottom)} The ratio between the $\gamma$-ray flux calculated using the 0.75$^\circ$ angular cut and the ratio utilizing our default 0.5$^\circ$  angular cut. The red upper limit corresponds to a bin where 1 photon is recorded using the larger angular cut, and 0 photons are recorded in the smaller cut. We note that the high-energy $\gamma$-ray flux remains almost identical, while the $\gamma$-ray flux around $\sim$10~GeV increases by only $\sim$20\%. These results indicate that our chosen ROI does not significantly affect our analysis.}
\label{fig:app_angularcuts}
\end{figure}

However, this angular cut also eliminates some true solar $\gamma$-rays with relatively poor angular reconstructions. The typical point-spread function for $\gamma$-ray events exceeding 10~GeV is 0.1$^\circ$ for front-converting events and 0.3$^\circ$ for back-converting events. However, this error is non-Gaussian, and the 95\% containment radius for back-converting events reaches nearly 1$^\circ$.  In Figure~\ref{fig:app_angularcuts}, we show the $\gamma$-ray spectrum calculated using a larger 0.75$^\circ$ ROI. We find that the solar $\gamma$-ray flux is only slightly affected, and the missing photons preferentially inhabit the lowest energy bins, which are not central to our main analyses. We find no trends indicating that the amplitude of these changes is larger in data taken either during or after solar minimum. This confirms that our analysis is resilient to the region of interest chosen for the analysis. 

\section{Tests for Instrumental Artifacts}
\label{appendix:phidependence}

Fermi's solar panels are fixed and remain oriented towards the Sun. Thus, the Sun occupies a unique position in the Fermi-LAT instrumental phase space. The vast majority of solar events are recorded within 3$^\circ$ of $\phi$~=~0, the coordinate that describes the direction normal to the plane of the solar panels. In~\citep{paper}, we perform two tests to examine the effect of this unique $\phi$-distribution on the energy spectrum of solar events. We first test any $\phi$-dependent feature that is already accounted for in the instrumental response functions, but smeared over in standard analyses (which assume that the exposure is $\phi$-independent). In particular, we grid the instrumental effective area over small portions of the $\phi$ parameter space using the hidden {\tt phibins} parameter in {\tt gtltcube}. This produces only a 4\% change in the overall Fermi-LAT exposure, and a $<$1\% change in the energy dependence of the Fermi-LAT exposure. In the default analysis shown in this paper, we have corrected for this effect by using 10 bins in instrumental $\phi$-space to calculate the Fermi-LAT effective area. We do not repeat the study of alternative $\phi$-binning options conducted in Ref.~\citep{paper}, as the impact of these choices is negligible. Moreover, the results do not depend on the specific ROI or energy cuts employed in each analysis, and thus the results are identical. 

\begin{figure*}[tbp]
\centering
\includegraphics[width=1.0\textwidth]{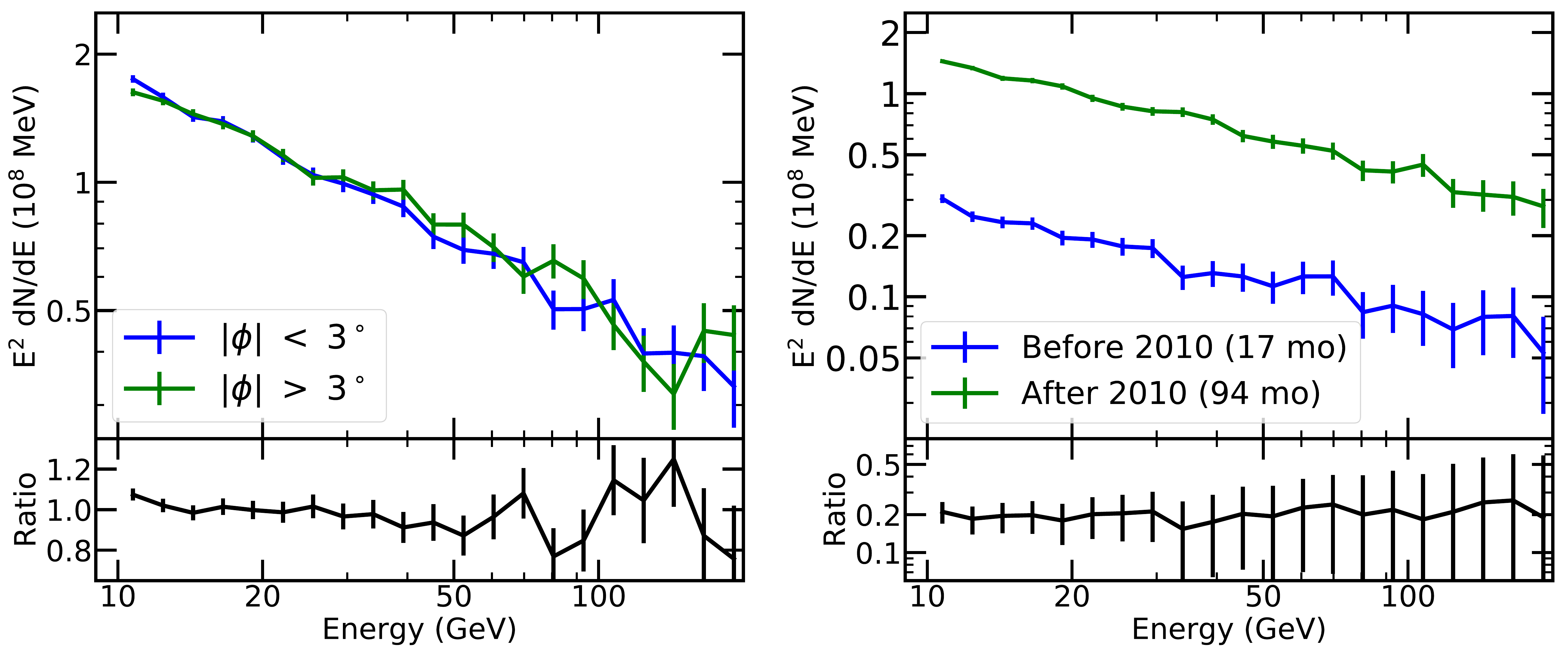}
\caption{ {\bf (Left top)} The energy spectrum of $\gamma$-rays above 10~GeV observed at least 5$^\circ$ from the solar position, but located within 3$^\circ$ of $\phi$~=~0 in Fermi-LAT instrumental coordinates (blue), compared to a ``partner" spectrum composed of photons that are nearby in {\it gcrs} coordinates but observed in instrumental coordinates exceeding 3$^\circ$ from $\phi$~=~0 (green).  {\bf (Left bottom)} The ratio between these datasets, which show that the spectral reconstruction in different regions of $\phi$-space is compatible with statistical errors. {\bf (Right top)} - The $\phi$~$<$~3$^\circ$ events shown in the left-panel, broken down into photons observed before (blue) or after (green) January 1, 2010. The difference in overall amplitude is explained by the much longer observation time (94 months vs 17 months) after 2010. {\bf (Right bottom)} - The ratio between these datasets, the $\sim$0.18 average of the data is compatible with the difference in exposures, and no spectral features are observed.}
\label{fig:app_phidependence}
\end{figure*}

The second test estimates the effect of ``unknown" systematic issues in the energy-reconstruction of events recorded near $\phi$~=~0 -- those which are not accounted for in the Fermi-LAT instrumental response functions. To examine this potential effect, we first extract every $\gamma$-ray event exceeding 10~GeV that is located at least 5$^\circ$ from the contemporaneous solar position (to avoid signal events), but was recorded at an instrumental $\phi$-angle within 3$^\circ$ of $\phi$~=~0. For each $\gamma$-ray in this set, we randomly select another ``partner" photon above 10~GeV that was observed at a similar position in {\it gcrs} coordinates, but at a $\phi$-value exceeding 3$^\circ$ from $\phi$~=~0. These partner photons are often found within 0.1$^\circ$ of the original $\gamma$-ray in {\it gcrs} coordinates, implying that their average $\gamma$-ray emission spectrum should be identical. Additionally, we note (from the previous test) that there are no known spectral features in the Fermi-LAT effective area near $\phi$~=~0. Thus, the count-spectra of the $\phi$~$\approx$~0 and ``partner" events should be the same. Additionally, in~Ref.~\citep{paper}, we confirm that this analysis technique does not induce any spectral features due to its hierarchical event selection procedure.

In Figure~\ref{fig:app_phidependence} (left) we show a spectral comparison between $\phi$~$\approx$~0 and ``partner'' events, finding no statistically significant spectral features that differentiate these datasets. This indicates that neither the $\gamma$-ray spectral dip nor the high $\gamma$-ray flux above 100~GeV are due to systematic artifacts relating to the instrumental $\phi$-dependence of the Fermi-LAT reconstruction. On the right, we show the spectrum of events observed near $\phi$~=~0 divided into two datasets corresponding to events observed before and after January 1, 2010. We find no distinguishing features between these spectra (except for the overall normalization that relates to the longer exposure after January 1, 2010). The combination of these observations argues against any instrumental systematics in the event reconstruction of solar events. We note that in Ref.~\citep{paper} we do find a systematic issue with an amplitude that peaks at approximately 10\% between $\sim$5--10~GeV for events observed near $\phi$~$\approx$~0 and partner events. We have found this miscalibration to be due to an increase in the misidentified cosmic-ray background near $\phi$~=~0. Our analysis here finds no evidence that this issue persists above 10~GeV.

\section{Background Estimation Using ``Fake Suns"}
\label{appendix:fakesuns}

\begin{figure*}[tbp]
\centering
\includegraphics[width=1.0\textwidth]{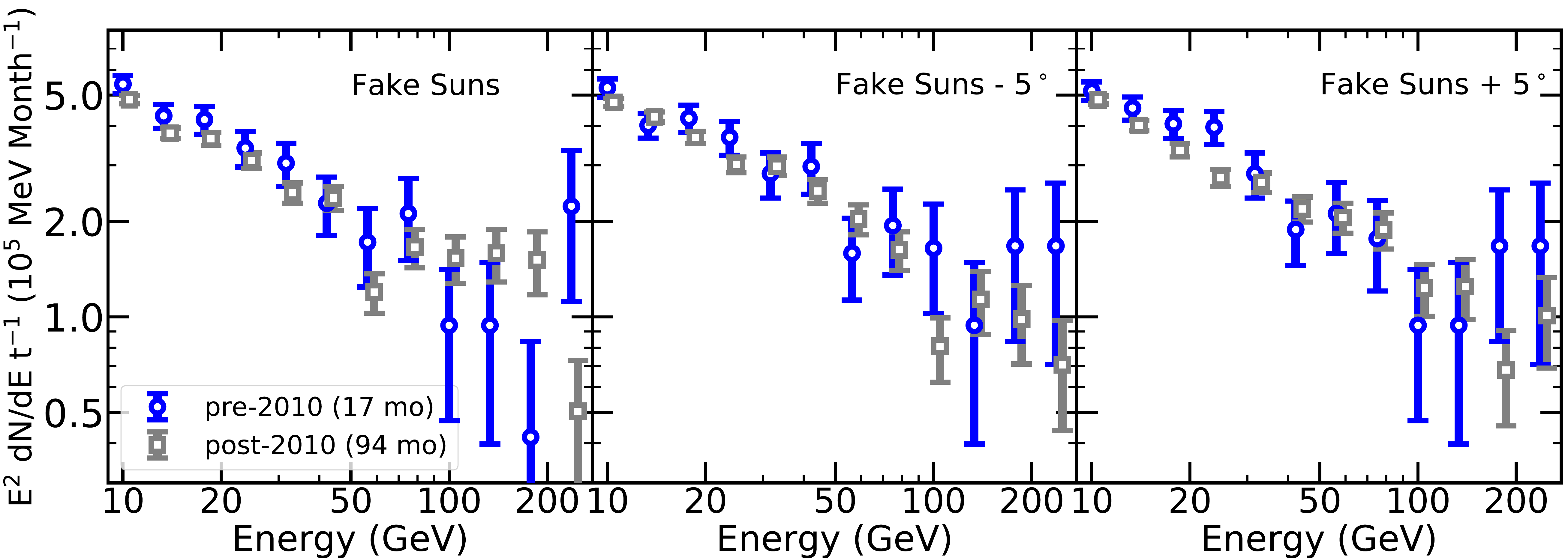}
\caption{The $\gamma$-ray emission from a stack of 355 ``fake Sun" positions that trail the Sun by $n$-days, where 5$<$n$<$360 (left). We find no spectral differences during and after the 2010 event selection cut, nor do we find evidence for a spectral-dip between $\sim$30--50~GeV. We repeat the process for 355 ``fake sun" positions that again trail the Sun by 5--360 days, but are shifted in Galactic longitude by either -5$^\circ$ (middle) or +5$^\circ$ (right). These cuts show no unexpected behavior. Because it is difficult to calculate the exposure for each fake Sun, the count spectrum is divided by the time-range in months, rather than the exposure. This approximately normalizes the data during and after solar minimum.}
\label{fig:app_fakesuns}
\end{figure*}

As in Ref.~\citep{paper}, we now estimate the diffuse $\gamma$-ray background and spectrum at the solar position by using an ensemble of ``fake Suns" that trail the solar position, but move through the same right ascensions and declinations. We select daily fake Suns that trail the real Sun by 5-360 days. We remove closer fake Suns due to contamination from solar inverse-Compton scattering. Due to the computational difficulty in obtaining the effective area for each fake Sun, we plot our results in terms of the total E$^2$~dN/dE counts, without accounting for the effective area. In the previous Appendix, we found that there are no unique features in the spectral or temporal evolution of the effective area at the true solar position, indicating that the spectrum of ``fake Suns" should be similar. 

In Figure~\ref{fig:app_fakesuns} we show that there are no significant spectral features at the position of ``fake Suns'' before or after Jan. 1, 2010. Moreover, we do not find any evidence of a spectral dip between $\sim$30--50~GeV. We find one low bin at 60~GeV (a higher energy than the dip found for the actual Sun), which is locally significant at $\sim$2.5$\sigma$, but is not globally significant. 

We stress that the diffuse flux shown here contributes negligibly to the solar $\gamma$-ray flux, which is a factor of $\sim$100~brighter at high energies. In particular, the 350 fake Suns produce 122 $\gamma$-rays above 100~GeV, indicating that the expected diffuse contribution to the $>$100~GeV $\gamma$-ray flux is $\sim$0.26 events over nine years, after accounting for the 25\% of the ROI that is blocked by the Sun itself. Thus, the lack of spectral features at fake Sun positions argues against any systematic effect stemming from the analyses utilized in the paper.

\section{Pass 6 and Pass 7 Data}
\label{appendix:pass6pass7}

\begin{table}[t]
\vskip 3mm 
\begin{tabular}{| c | c | c | c | c | c | c |}
\hline
Time (UTC) & E (GeV) & R.A. & Dec & $\theta_\odot$ & Event Class\\ \hline
2009-08-21 10:14:32 & 179.1 & 150.59 & 12.08 & 0.075$^\circ$& UltraClean \\
\hline
2014-10-31 14:12:01 & 201.1 & 215.78 & -14.28 & 0.289$^\circ$& Source\\
\hline

\end{tabular}
\caption{Event Information for P7V6\_Source events $>$100~GeV that were recorded within 0.5$^\circ$ of the solar position, but were not reconstructed as $\gamma$-rays by the Pass 8 event reconstruction utilized in the main text. We find two events. One was observed during solar minimum, while one was observed near solar maximum.  We note that Pass 6 and Pass 7 events do not cover the full time-range of our analysis, as event reconstruction ended in August, 2011 and June, 2015, respectively. The first event was also found in Pass 6, while the second lies outside the Pass 6 time window.}
\label{app:tab:alternativecuts}
\end{table}

Due to the small number of solar $\gamma$-rays above 100~GeV, it is worth examining events that did not make our standard analysis cuts. In the main text, we unmasked the region $|$b$|<~$5$^\circ$, finding two additional events above 100~GeV that were observed during solar minimum. Here, we examine events that did not pass the Pass 8 Source event cuts, but were recorded as source-class events in Pass 6 or Pass 7. 

In Table~\ref{app:tab:alternativecuts}, we list the two events recorded in Pass 7 data. One event was observed during solar minimum, while one was observed near solar maximum. Including these two events in our analysis would slightly increase the statistical significance of temporal variability, though this is hard to quantitatively measure because the effective area differs between Pass 7 and Pass 8. We note that the first event was also recorded in Pass 6, while the second event was recorded after Pass 6 reconstruction ended. We additionally note that Pass 7 reconstructions ended in June 2015, before the end of our analysis.

\section{Calculation of the Maximum Solar Flux}
\label{appendix:maximum}

Here, we describe our calculation of the maximum solar disk flux, assuming a solar disk emission mechanism identical to that posited by SSG (and roughly described in Equation~\ref{eq:1}). We first note that the maximum flux from the solar disk can be roughly approximated by taking the SSG model, which assumes f$_{\rm sur}$~=~1.0, f$_{\rm turn}$~=~1 and f$_{\rm int}$~=~0.005 and multiplying by 200. This would describe a scenario where every incoming $\gamma$-ray that encounters the solar surface has its direction reversed by solar magnetic fields, and then subsequently converts the entirety of its cosmic-ray energy to $\gamma$-rays (and neutrinos) that are not further attenuated by the solar photosphere. However, this scenario is not realizable, in part because the f$_{\rm int}$~ is calculated utilizing an optical depth that depends sensitively on the incoming cosmic-ray angle. For cosmic-rays that are incident at angles perpendicular to the solar surface, the optical depth already exceeds 1 in the SSG model, indicating that the $\gamma$-ray emission from these cosmic-rays cannot be boosted further. 

Here, we produce a very conservative upper limit on the $\gamma$-ray flux from cosmic-ray interactions in the solar photosphere, which we have shown in Figure~\ref{fig:spectrumratio} and Figure~\ref{fig:app_angularcuts}. This calculation closely follows the methods of SSG (and provides similar results). However, we detail this calculation for three reasons: (1) to provide clarity concerning the limit shown in this text, (2) to update the interstellar cosmic-ray fluxes and cosmic-ray shower physics to better reflect current data, (3) to formally extend the calculation to TeV energies (SSG ended at a $\gamma$-ray energy of 5~GeV, although an extrapolation is possible). 

\begin{figure*}[tbp]
\centering
\includegraphics[width=0.48\textwidth]{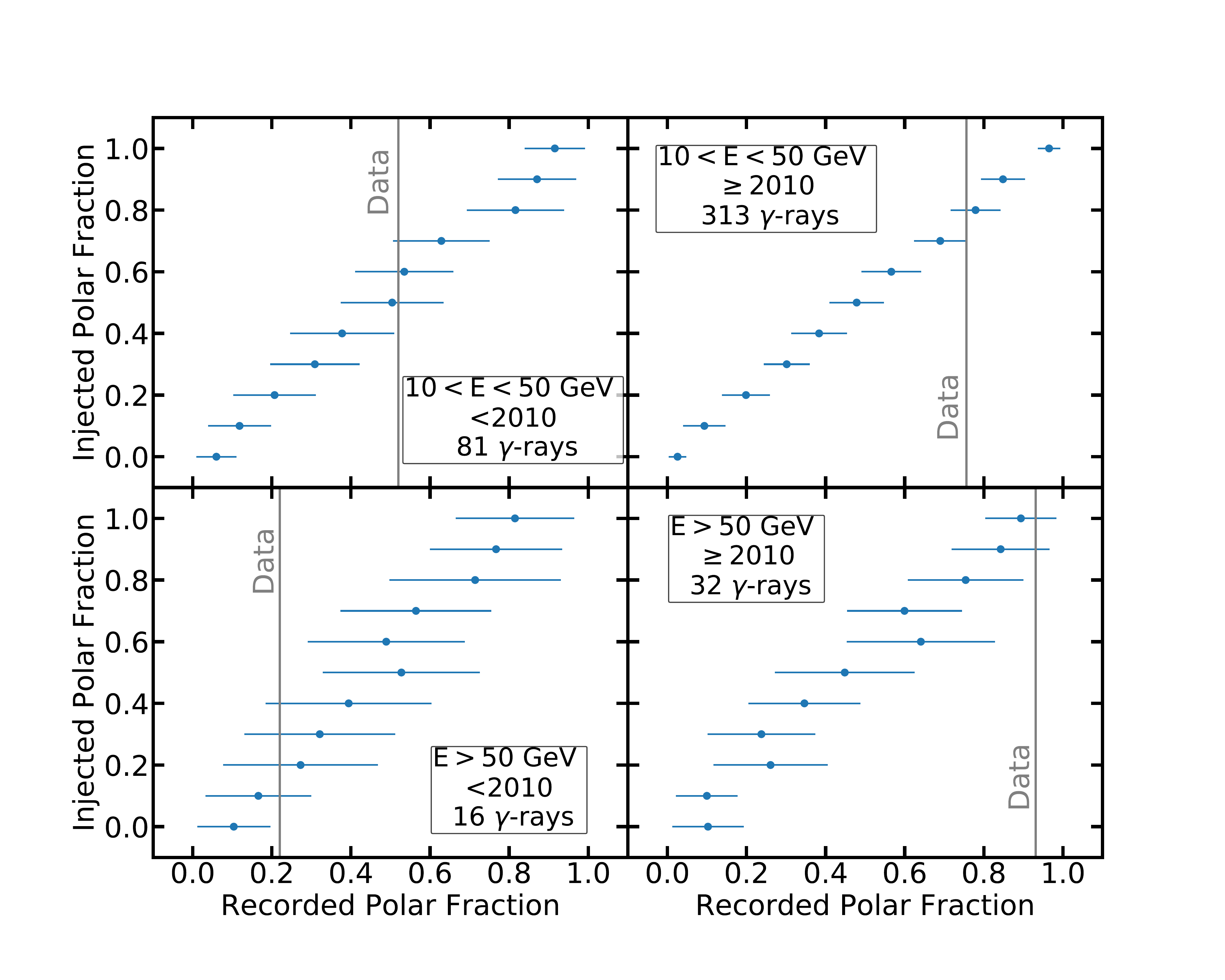}
\includegraphics[width=0.48\textwidth]{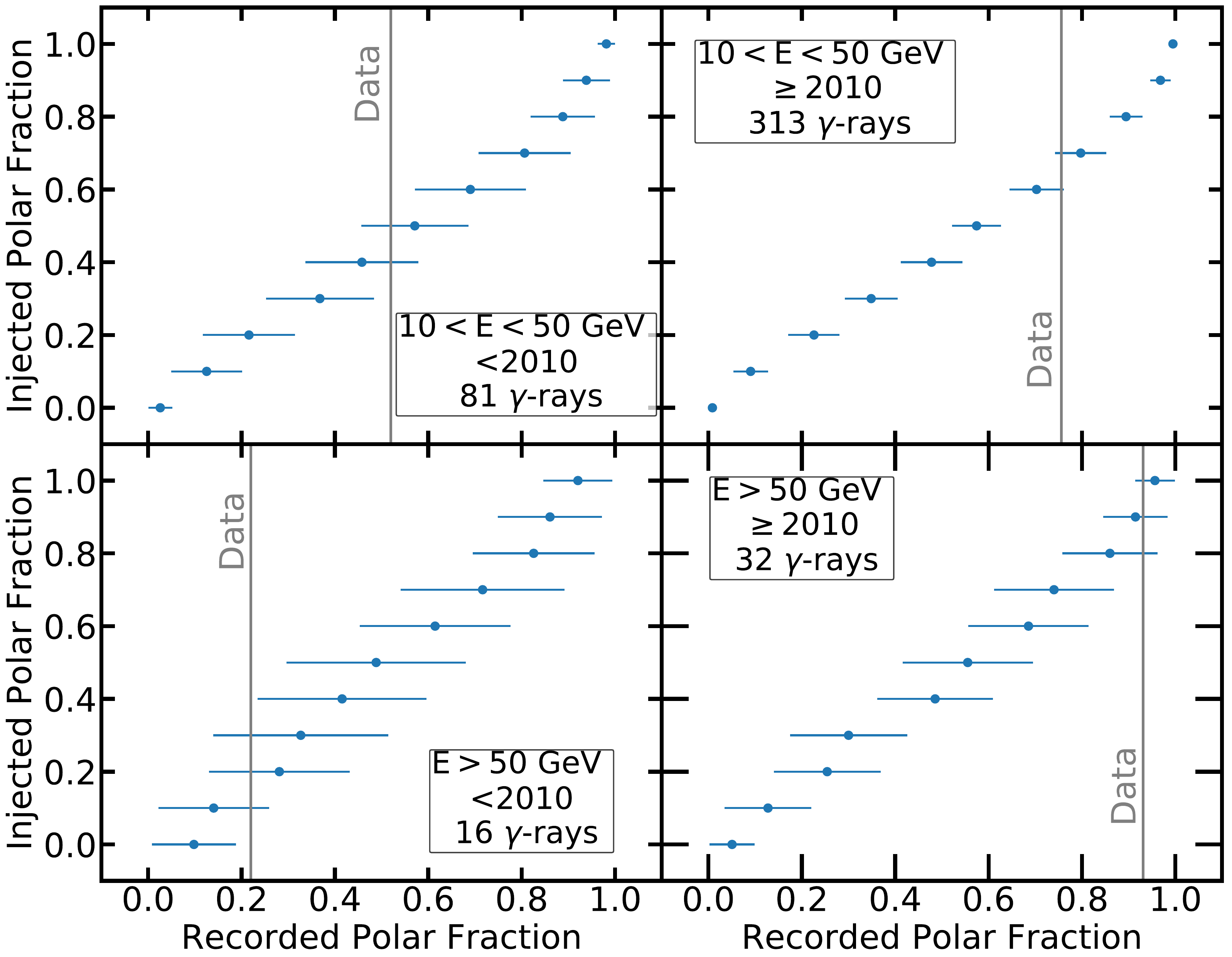}
\caption{The distribution of recorded polar fluxes produced by mock realizations of solar $\gamma$-ray emission with a given ``true'' polar flux fraction. Fractions are expressed such that the polar and equatorial fluxes sum to unity. Simulated data are taken from 10 Monte Carlo simulations for each injected polar fraction, and the data is divided into emission before and after January 1, 2010, as well as below and above 50~GeV, as in the main text. We utilize two extremal models for the true morphology of solar $\gamma$-rays. The first (left) includes polar (equatorial) components that emit isotropically outside (inside) the $|$T$_y|$~$<$~0.108$^\circ$ cut employed in our analysis. The second (right), includes a polar component that emits only from the solar poles, and an equatorial component that emits only along the equatorial plane. In both cases, we find that the temporal shift from equatorial to polar emission is statistically significant above 50~GeV, but is only marginally significant at lower energies.}
\label{fig:app:morphologytest}
\end{figure*}

We include an initial flux of both cosmic-ray protons and helium, which provides an important enhancement due to the larger nuclear cross-section~\citep{Zhou:2016ljf}. We conservatively assume that all incoming cosmic-rays are efficiently turned by solar magnetic fields before any hadronic interactions begin, and thus no cosmic-ray energy is lost in the turning process. 

We then set up a GEANT4 simulation~\citep{Agostinelli:2002hh} to calculate the outgoing $\gamma$-ray flux from an incoming cosmic-ray flux encountering photospheres of varying optical depth, and fit the optical depth to maximize the outgoing $\gamma$-ray flux. Thus, this calculation assumes that cosmic-rays encountering the solar surface at any pitch angle encounter the optimal amount of solar material needed to maximize the outgoing $\gamma$-ray flux. Furthermore, we note that this simulation naturally takes into account effects such as $\gamma$-ray production from secondary cosmic-rays, and $\gamma$-ray absorption by the remaining photosphere. We find the optimal grammage to be $\sim$130~g~cm$^{-2}$, which translates to a proton optical depth of $\tau$~$\sim$~2.3. We find that this (extremely conservative) calculation produces a maximum upper limit that exceeds the standard SSG prediction by a factor of $\sim$100, slightly smaller than the factor of 200 enhancement unrealistically calculated by naively using the SSG prediction and setting f$_{\rm int}$~=~1.0.

\section{Morphological Modeling}
\label{appendix:morphologymodeling}

In Figure~\ref{fig:helioprojective}, we used the reconstructed positions of $\gamma$-rays in helioprojective coordinates to determine the fraction of the solar $\gamma$-ray emission produced by polar and equatorial regions across the solar surface. In this appendix, we describe the process in more detail and verify our analysis utilizing mock signals with known polar and equatorial components. 

Our morphological analysis proceeds as follows. We decompose the solar surface into an ``equatorial'' region defined by $|$T$_y|$~$<$~0.108$^\circ$, and a ``polar'' region that fills the remaining solar surface. These regions occupy an equal area of the solar surface (though we note that our analysis does not require the area in each region to be equivalent). For computational simplicity, we randomly populate each emission region with 3000 ``point'' sources of equivalent luminosity to approximate a diffuse emission component. For every $\gamma$-ray in our data sample, we utilize information regarding its energy, psf-class and observed $\theta$-angle to calculate the double-King function PSF described in the Fermi-LAT Cicerone~\footnote{https://fermi.gsfc.nasa.gov/ssc/data/analysis/documentation/Cicerone/\\Cicerone\_LAT\_IRFs\/IRF\_PSF.html}. Utilizing this information, we calculate the relative probability that a each observed $\gamma$-ray came from all 6000 point source locations.

\begin{table*}[!htb]
\vspace{0.08cm}
\centering{Model Fluxes with the Inclusion of Limb and Background Emission Component}
\vspace{0.02cm}
    \begin{minipage}{.5\linewidth}
      \centering Before January 1, 2010, 10~GeV $<$ E~$<$~50~GeV \\
      \centering (Fluxes in 10$^{-6}$~MeV~cm$^{-2}$~s$^{-1}$)\\
      \vspace{0.05cm}
\begin{tabular}{| c | c | c |}
\hline\hline
Component & Posterior Fit & Best-Fit  \\
\hline
Polar & 8.0~$\pm$~2.6 & 9.2 \\
Equatorial & 12.9~$\pm$~2.9 & 13.6 \\
Limb & 2.9~$\pm$1.9 & 1.3 \\
Background & 1.5~$\pm$~1.0 & 1.2\\
\hline
\end{tabular}
    \end{minipage}%
    \begin{minipage}{.5\linewidth}
      \vspace{0.3cm}
      \centering After January 1, 2010, 10~GeV $<$ E~$<$~50~GeV\\
      \centering (Fluxes in 10$^{-6}$~MeV~cm$^{-2}$~s$^{-1}$)\\
      \vspace{0.05cm}
\begin{tabular}{| c | c | c | c |}
\hline\hline
Component & Posterior Fit & Best-Fit  \\
\hline
Polar & 8.4~$\pm$~1.4 & 8.5 \\
Equatorial & 3.6~$\pm$~1.0 & 3.6 \\
Limb & 5.9~$\pm$1.3 & 5.9 \\
Background & 0.7~$\pm$~0.4 & 0.5\\
\hline
\end{tabular}
\vspace{0.4cm}    
    \end{minipage} 
\vspace{0.4cm}    
    \begin{minipage}{.5\linewidth}
      \centering Before January 1, 2010,  E~$>$~50~GeV\\
      \centering (Fluxes in 10$^{-6}$~MeV~cm$^{-2}$~s$^{-1}$)\\
      \vspace{0.05cm}
\begin{tabular}{| c | c | c | c |}
\hline\hline
Component & Posterior Fit & Best-Fit  \\
\hline
Polar & 3.9~$\pm$~2.9 & 3.5 \\
Equatorial & 16.8~$\pm$~3.3 & 17.6 \\
Limb & 3.3~$\pm$2.3 & 1.5 \\
Background & 1.7~$\pm$~1.4 & 3.1\\
\hline
\end{tabular}

    \end{minipage}%
    \begin{minipage}{.5\linewidth}
      \centering After January 1, 2010,  E~$>$~50~GeV\\
      \centering (Fluxes in 10$^{-6}$~MeV~cm$^{-2}$~s$^{-1}$)\\
      \vspace{0.05cm}
\begin{tabular}{| c | c | c | c |}
\hline\hline
Component & Posterior Fit & Best-Fit  \\
\hline
Polar & 4.4~$\pm$~1.1 & 5.5 \\
Equatorial & 0.6~$\pm$~0.5 & 0.4 \\
Limb & 1.5~$\pm$1.0 & 0.9 \\
Background & 0.3~$\pm$~0.3 & 0.0\\
\hline
\end{tabular}

\vspace{-0.1cm}
    \end{minipage} 
\vspace{-0.3cm}
\caption{The decomposition of the observed solar $\gamma$-ray flux into a model featuring four components, including the equatorial and polar components utilized in the main text, as well as a limb component and background component described in the Appendix. The energy and temporal cuts are identical to those employed throughout the main text. In addition to the posterior fit distribution from our Bayesian analysis, we show the best-fit value from a scan of our posterior chain. We note three key results: (1) the posterior distribution and best-fit values are in good agreement, (2) the background component contributes negligibly to the total emission, validating our choice to ignore this component in the main analysis, (3) the temporal evolution of the equatorial component is qualitatively unaffected (and quantitatively enhanced) by the inclusion of additional model components.}
\label{app:tab:models_4component}
\end{table*}

We then convolve the relative diffuse contributions over all observed $\gamma$-rays to produce a best-fit model of the total flux from the Sun's polar and equatorial regions. In particular, we maximize the function $\prod_{i}~(\alpha_{pol}P_{pol, i} + \alpha_{eq}P_{eq, i})$, where P$_{eq}$ and P$_{pol}$ are the probabilities that the ensemble of points in each distribution can produce an observed event $i$, while $\alpha_{eq}$ and $\alpha_{pol}$ are the global parameters that describe the best-fitting fraction of the total polar and equatorial emission. The product is taken over all observed $\gamma$-rays. Because these two components are assumed to produce the entire solar emission, we constrain $\alpha_{eq}$~=~1-$\alpha_{pol}$. We utilize a Multinest scan, beginning with a flat Bayesian prior of 0$<\alpha_{pol}<$1.
 
To determine the accuracy of this procedure, we produce a Monte Carlo analysis using mock photon samples in each energy and temporal bin utilized in our main analysis. Specifically, for each energy and temporal cut, we produce a fake sample of photons with a photon count that is identical to observations, a photon spectrum which falls as E$^{-2}$ within each bin, and a $\theta$-distribution that is biased as $\theta$, and a random selection of photon counts from each PSF-class. This approximately matches the distribution of $\gamma$-rays expected for real events, but is somewhat conservative both because the $\theta$-distribution is broader, and the PSF-distribution is slightly biased towards poorly reconstructed events. For each model event, we produce a true solar $\gamma$-ray direction, and then choose a random observed location that depends on the photons individual PSF, as calculated above.   
 
For each energy and temporal bin, we produce 10 Monte Carlo realizations for 11 different injected distributions of polar and equatorial emission, which we scan in 10\% steps from 100\% polar emission to 100\% equatorial emission. We note three important facts: (1) this ratio is imposed on the relative flux, and not on the observed photon count. The Poisson fluctuations in the relative photon count (e.g. for the 16 $\gamma$-rays that were recorded during solar minimum and above 50~GeV) actually dominates the error budget in our reconstruction, (2) while we utilize models for the polar and equatorial emission components that are isotropic in their respective ROIs, we do not know how the ``true'' emission morphology is biased within each component. We assume two extremal cases. In the first, we assume that the true underlying $\gamma$-ray morphology reflects our model, that is, the true polar emission is isotropic across the solar disk in regions with $|$T$_y|$~$>$~0.108$^\circ$, while the ``equatorial'' emission component is isotropic for smaller values of T$_y$. This model is conservative, because there is no biasing of the emission component away from its boundaries. In the second model, we assume an extremal scenario where the true ``polar'' emission is produced only at the north and south poles, while the true ``equatorial'' emission is produced only along the solar equatorial plane.

\begin{table*}[!htb]
\vspace{0.08cm}
\center{Model Fluxes with the Inclusion of Independent North Polar and South Polar Components} 
\vspace{0.2cm}
    \begin{minipage}{.5\linewidth}
      \centering Before January 1, 2010, 10~GeV $<$ E~$<$~50~GeV \\
      \centering (Fluxes in 10$^{-6}$~MeV~cm$^{-2}$~s$^{-1}$)\\
      \vspace{0.05cm}
\begin{tabular}{| c | c | c |}
\hline\hline
Component & Posterior Fit & Best-Fit  \\
\hline
Equatorial & 14.1~$\pm$~2.5 & 13.7 \\
North Polar & 5.1~$\pm$~1.7 & 3.4 \\
South Polar & 6.1~$\pm$~1.9 & 8.2 \\
\hline
\end{tabular}

    \end{minipage}%
    \begin{minipage}{.5\linewidth}
     \vspace{0.3cm}
      \centering After January 1, 2010, 10~GeV $<$ E~$<$~50~GeV \\
      \centering (Fluxes in 10$^{-6}$~MeV~cm$^{-2}$~s$^{-1}$)\\
      \vspace{0.05cm}
\begin{tabular}{| c | c | c |}
\hline\hline
Component & Posterior Fit & Best-Fit  \\
\hline
Equatorial & 4.9~$\pm$~1.0 & 4.0 \\
North Polar & 6.8~$\pm$~0.7 & 6.7 \\
South Polar & 6.8~$\pm$~0.8 & 7.8 \\
\hline
\end{tabular}

\vspace{0.4cm}    
    \end{minipage} 
\vspace{0.4cm}    
    \begin{minipage}{.5\linewidth}
      \centering Before January 1, 2010, E~$>$~50~GeV \\
      \centering (Fluxes in 10$^{-6}$~MeV~cm$^{-2}$~s$^{-1}$)\\
      \vspace{0.05cm}
\begin{tabular}{| c | c | c |}
\hline\hline
Component & Posterior Fit & Best-Fit  \\
\hline
Equatorial & 15.7~$\pm$~3.8 & 18.5 \\
North Polar & 8.0~$\pm$~3.6 & 4.1 \\
South Polar & 1.8~$\pm$~1.7 & 2.9 \\
\hline
\end{tabular}

    \end{minipage}%
    \begin{minipage}{.5\linewidth}
      \centering After January 1, 2010, E~$>$~50~GeV \\
      \centering (Fluxes in 10$^{-6}$~MeV~cm$^{-2}$~s$^{-1}$)\\
      \vspace{0.05cm}
\begin{tabular}{| c | c | c |}
\hline\hline
Component & Posterior Fit & Best-Fit  \\
\hline
Equatorial & 0.7~$\pm$~0.6 & 0.9 \\
North Polar & 3.3~$\pm$~0.7 & 2.2 \\
South Polar & 3.0~$\pm$~0.7 & 2.9 \\
\hline
\end{tabular}

\vspace{-0.1cm}
    \end{minipage} 
\vspace{-0.3cm}
\caption{Same as Table~\ref{app:tab:models_4component}, for an alternative morphological model that includes the default equatorial component, but divides the polar component into two independent northern and southern components. We find that this model is qualitatively consistent with our default analysis. In particular, it finds a significant decrease in the equatorial flux after solar minimum -- and finds that the amplitude of this shift increases significantly at high energies. Additionally, the model finds no indication of time or spectral variations in either polar component.  We find no statistically significant evidence for a north-south equatorial split in any of the temporal and energy bins shown here. While a north-south split in data taken before January 1, 2010 at energies above 50~GeV in Figure~\ref{fig:helioprojective} is striking to the eye, we find the statistical significance of this feature is below 2$\sigma$.}
\label{app:tab:models_splitpolar}
\end{table*}

In Figure~\ref{fig:app:morphologytest}, we show the results from our Monte Carlo simulations, finding that our model can distinguish between polar and equatorial emission with a high degree of statistical significance. In general, we find uncertainties of $\sim$20\% in the reconstructed polar fraction for photons above 50~GeV, and uncertainties of approximately 10\% for photons from 10-50~GeV. The smaller uncertainties at lower energies stem from the much larger photon count in this energy range, which decreases the Poisson fluctuations in the number of counts observed from the true polar and equatorial emission components. We show (as a vertical line), the reconstructed polar fraction for the observed solar data --- finding that no single ``true'' underlying polar fraction can explain the observed data taken both during and after the solar minimum. This verifies the key result shown in the main text -- that the transition from a primarily equatorial $\gamma$-ray morphology during solar minimum, to a primarily polar $\gamma$-ray morphology during the remaining solar cycle -- is real. 

We do note that the uncertainties in the reconstructed polar and equatorial fluxes exceed the statistical uncertainties for a single realization of the data (and reported in the main text). This is primarily due to the large Poissonian errors present in the low-photon count regime. Continued observations by the Fermi-LAT (in particular during the next solar minimum) will significantly decrease these errors, allowing us to more accurately probe the underlying morphology. Finally, we note that while our first model of the underlying photon emission morphology (left) is relatively bias free, the extremal emission model (right) shows a slight (5\%) bias towards polar emission. This effect is negligible compared to the observed shifts in the flux ratio at both low and high energies. 

\section{Alternative Morphological Models}
\label{appendix:limbbackground}

In Figure~\ref{fig:helioprojective}, we showed that the solar disk $\gamma$-ray emission can be broken down into equatorial and polar components, and that the intensity of equatorial emission greatly decreases at the end of solar minimum, while the polar emission component remains relatively constant. However, we note that the choice to divide the data into ``equatorial" and ``polar" models was made heuristically, and additional morphological models may contribute to the observed $\gamma$-ray data.

Here, we test two additional, well-motivated models. Our first model includes both the equatorial and polar components defined in the main text, but adds both a ``limb" component, which emits photons with radial symmetry at an angular distance of  $\theta_{Sun}$~$\sim$~0.26$^\circ$ from the solar center, and a ``background" component, which has a constant surface brightness over the remaining 0.5$^\circ$ ROI that is not blocked by the sun.  The limb model physically corresponds to emission produced when cosmic-rays directed towards Earth hit solar gas in an optically thin region, producing $\gamma$-ray emission beamed towards Earth. The ``background" model corresponds to $\gamma$-ray point sources and diffuse emission near the Sun. It also approximately accounts for the contribution of solar inverse-Compton scattering $\gamma$-rays. While these are not isotropic, they are suppressed by Klein-Nishina effects across the solar disk, and peak in intensity immediately outside the disk.

In Table~\ref{app:tab:models_4component}, we calculate the contribution of each emission component to the solar $\gamma$-ray flux, utilizing the standard temporal and energy cuts employed throughout this paper. We find that, even though all four emission components are allowed vary freely, the polar and equatorial emission components continue to produce the majority of the observed flux, while the limb and background components are subdominant. This validates our utilization of a polar and equatorial models in our default analysis. We note that the background component, in particular, is often compatible with 0, validating our choice to disregard the contribution of diffusion emission in the main text. We note that this result is also validated by our analysis of fake-Sun positions in Appendix~\ref{appendix:fakesuns}. Most importantly, we find that our quantitative conclusions regarding the temporal variation in the equatorial emission component and the steady-state nature of the polar emission remain consistent in this model. This provides evidence that our standard model is resilient to reasonable changes in the choice of morphological models. In fact, we find that the amplitude of the time-variation in the equatorial component increases in this alternative model, as some fraction of the post-2010 equatorial emission shifts into the limb component. Unfortunately, our statistical sample is not large enough to make any strong statement regarding this trend.

In Table~\ref{app:tab:models_splitpolar}, we show a second alternative morphological model, which includes the same equatorial component as our main analysis, but splits the polar emission into ``north'' and ``south'' components that are normalized independently. This model is motivated by the absence of $>$50~GeV $\gamma$-rays observed in the southern disk before January 1, 2010. We again find that this model qualitatively recovers all key conclusions from the main text, including the temporal and spectral variation in the equatorial component, and the steady-state nature of both polar components. While our model does indicate that the northern polar flux exceeds the southern polar flux above 50~GeV during solar minimum, we find that this is only statistically significant at $\sim$1.5$\sigma$ locally. Because we have tested several time and energy bins, we additionally take a trials factor. We conclude that there is no statistically significant feature that separates northern and southern polar emission in our model.

\begin{figure}[tbp]
	\centering
	\includegraphics[width=0.48\textwidth]{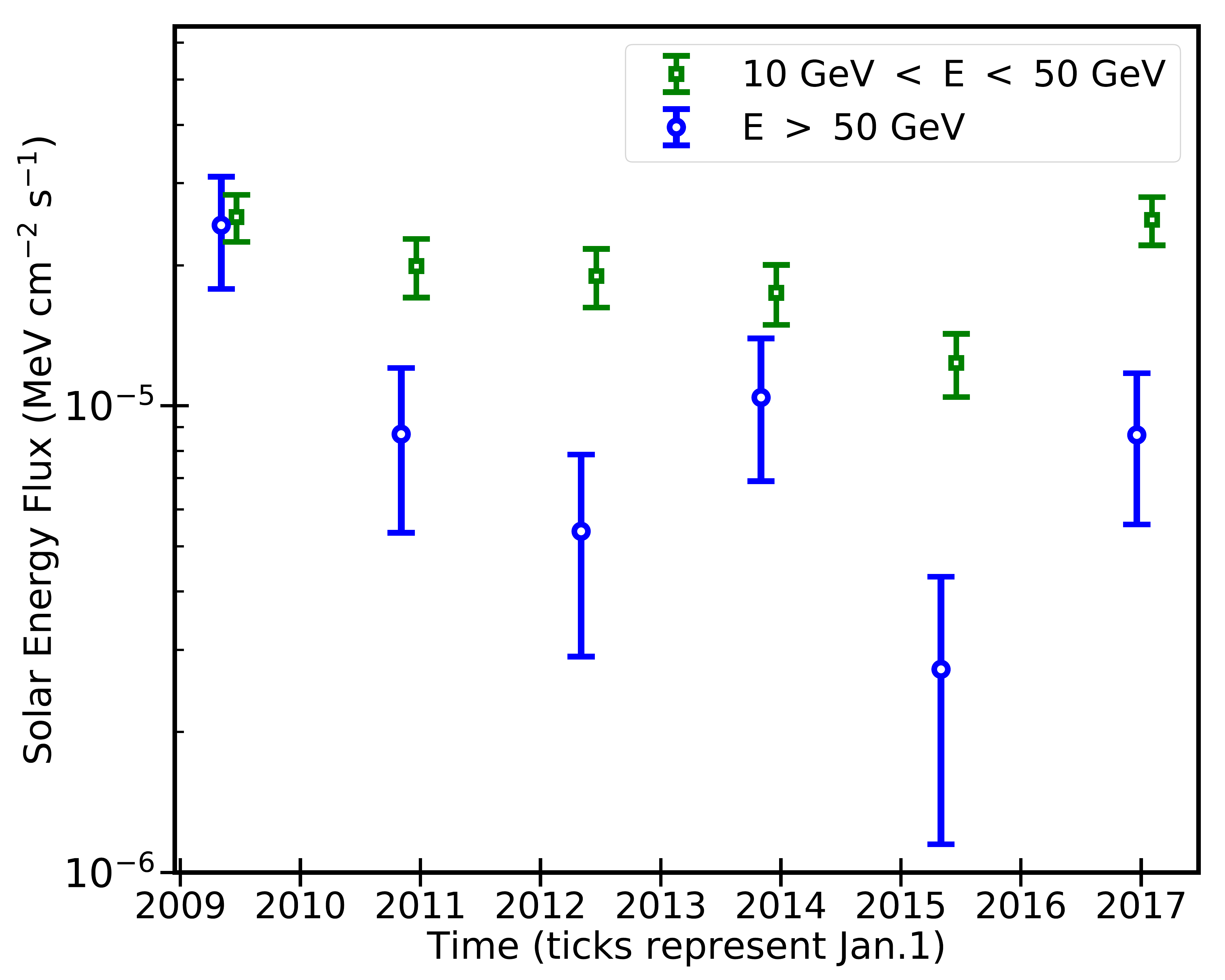}
	\caption{The $\gamma$-ray flux in our ROI binned into 1.5~yr increments spanning our analysis period (the last bin is slightly longer). Between 10---50~GeV, the $\gamma$-ray flux peaks during solar minimum, but declines both moderately and smoothly towards solar maximum, in agreement with~\citep{paper}. However, at higher energies, the $\gamma$-ray flux drops sharply from solar minimum to the remaining solar cycle.}
	\label{app:fig:timevariability}
\end{figure}

\section{Analysis Before and After the Heliospheric Flip}
\label{appendix:flip}

In our default analysis, we assumed that the intensity of the solar magnetic field dictates the amplitude, spectrum and morphology of the solar disk $\gamma$-ray flux. This choice is motivated by the peculiar time variation of solar $\gamma$-rays above 100~GeV. However, alternative temporal cuts are possible, and may be motivated by known solar processes. One mechanism capable of producing a temporal shift in the solar $\gamma$-ray flux and morphology is the orientation of the large-scale heliospheric magnetic field (HMF). The HMF follows a 22 year cycle, with polarity reversals every 11~years during the solar maximum. The HMF propagates throughout the solar system, carried by the heliospheric current sheet. In particular, Voyager observed polarity changes in the HMF in the outer solar system~\citep{2002JGRA..107.1410B}. 

\begin{figure}[tbp]
	\centering
	\includegraphics[width=0.48\textwidth]{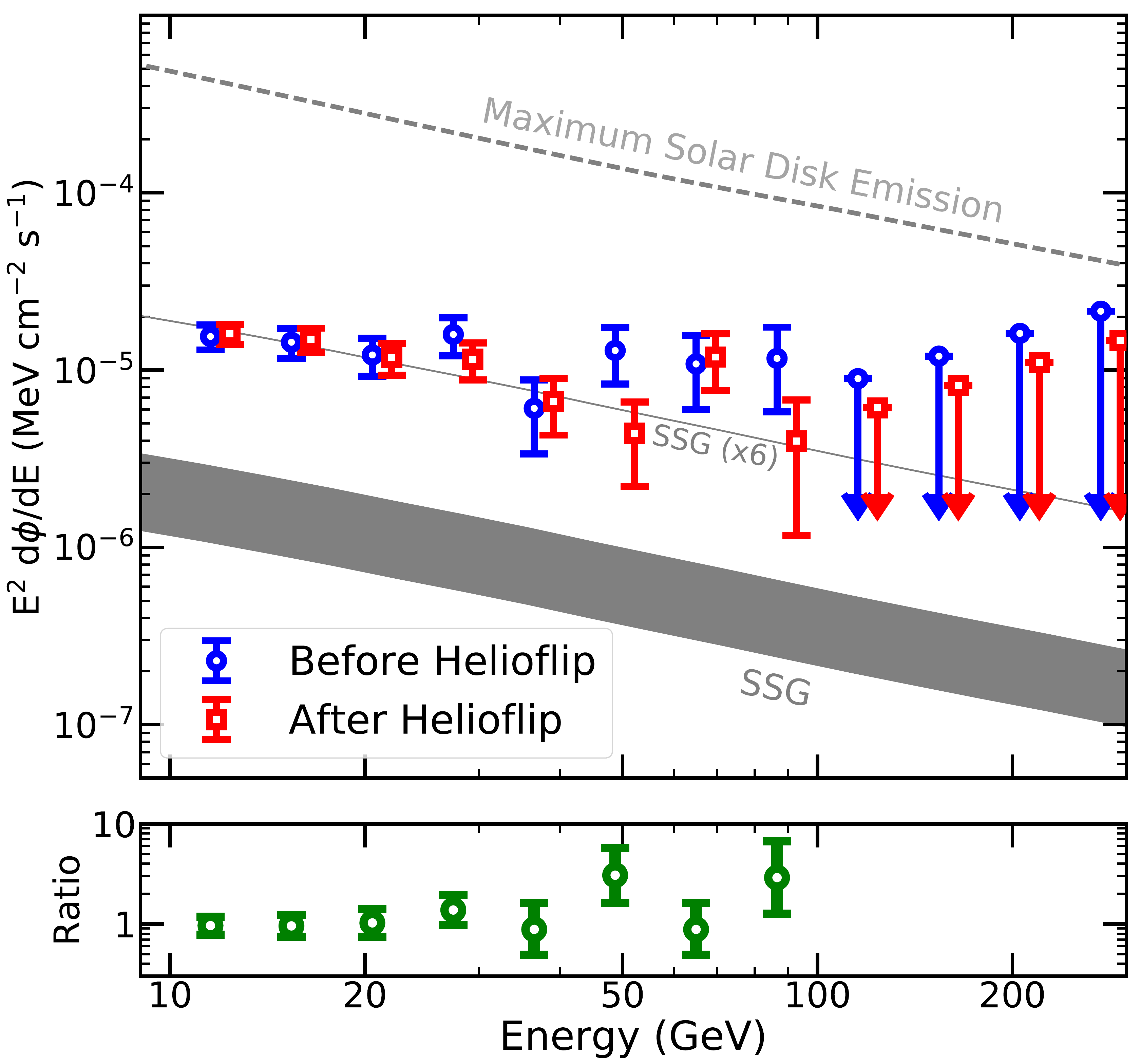}
	\caption{Same as Figure~\ref{fig:spectrumratio}, but using temporal cuts that isolate the emission before and after the heliospheric magnetic field flip. Because the flip is not instantaneous, we use data from January 1, 2010 -- August 10, 2012 before the flip, and January 15, 2014 --- November 5, 2017 after the flip. We find no evidence for spectral or intensity changes between these periods. We note that no events $>$100~GeV are observed in either period, rendering the ratio ill-defined.}
	\label{app:fig:spectrumheliospheric}
\end{figure}

During periods of positive (A$>$0) polarity, positively charged cosmic-rays primarily diffuse to Earth through polar regions, while during periods of negative (A$<$0) polarity, positive cosmic-rays primarily drift across the heliospheric current sheet, where particle diffusion is highly constrained~\citep{2012Ap&SS.339..223S}. At lower ($<$10~GeV) energies, the decreased propagation efficiency across the heliospheric current sheet results in increased solar modulation, which is observed at Earth by instruments like PAMELA and AMS-02~\citep{2012Ap&SS.339..223S, Cholis:2015gna}. 

\begin{figure*}[tbp]
\centering
\includegraphics[width=0.9\textwidth]{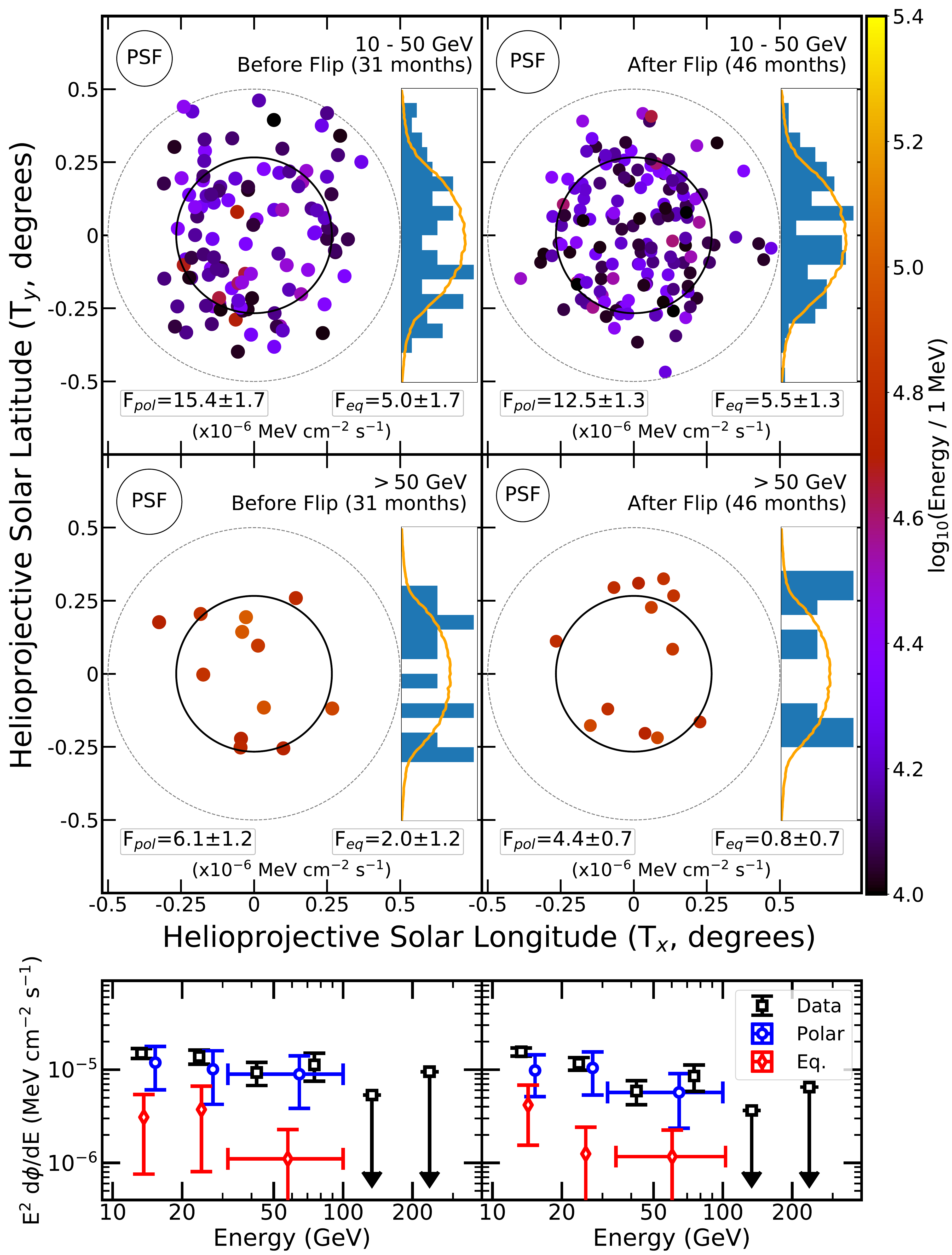}
\caption{Same as Figure~\ref{fig:helioprojective}, but utilizing temporal cuts that isolate the period before and after the Sun's heliospheric magnetic field flip (which occurred between approximately August 10, 2012 and January 15, 2014~\citep{2016ApJ...823L..15G}). The period during the Heliospheric magnetic field flip, as well as the period of solar minimum (before January 1, 2010) are excluded from the analysis to isolate contributions from the heliospheric magnetic field polarity reversal. We find no evidence for significant changes in the $\gamma$-ray morphology due to the global change in the heliospheric magnetic field polarity.}
\label{app:fig:helioprojective_helioflip}
\end{figure*}

While the strength of the HMF near Earth is insufficient to affect TeV cosmic-rays, it is worth considering whether this large-scale field could affect higher-energy cosmic-rays once they propagate closer to the Sun. In Figure~\ref{app:fig:timevariability}, we test this possibility by plotting the solar $\gamma$-ray flux in 1.5~yr increments. At both low and high energies, the $\gamma$-ray flux is largest at solar minimum and declines thereafter. However, the amplitude of these variations is not consistent. At low energies, the total $\gamma$-ray flux decreases moderately and smoothly -- and begins to rise as we again approach solar minimum. At high-energies the $\gamma$-ray flux drops precipitously between solar minimum and the remaining cycle. The high-energy flux outside of solar minimum appears constant, though we note that the statistical uncertainty becomes large.  

In Figure~\ref{app:fig:spectrumheliospheric}, we plot the $\gamma$-ray spectrum before and after the  HMF polarity flip. We first remove two periods: (1) data taken during the solar minimum, to eliminate contributions from the low magnetic field during this period, and (2) data taken during the polarity flip, when the polarity is poorly defined. While the exact beginning and end of the field reversal is poorly defined, we choose August 10, 2012 to January 15, 2014~\citep{2016ApJ...823L..15G}.  This leaves us with 31 (46) months of exposure in the before (after) the polarity flip.  While the statistical significance of this sample is poorer, we find no evidence for intensity or spectral variations before and after the polarity flip, indicating that it has a small effect on the $\gamma$-ray emission.

Finally, while the HMF does not significantly affect either the amplitude or spectrum of solar $\gamma$-rays, it may still affect its morphology by producing anisotropies in the incoming cosmic-ray flux. In Figure~\ref{app:fig:helioprojective_helioflip}, we show the morphology of solar $\gamma$-ray emission before and after the magnetic field flip, following the standard methodology to differentiate equatorial and polar emission. We find no evidence for any significant shift in the $\gamma$-ray emission morphology before and after the magnetic field flip. This argues against any large-scale magnetic field contribution to the morphology of the solar disk $\gamma$-ray flux. \newline

\section{Morphology of the 30--50~GeV Spectral Dip}
\label{appendix:polardip}

\begin{figure}[tbp]
	\centering
	\includegraphics[width=0.48\textwidth]{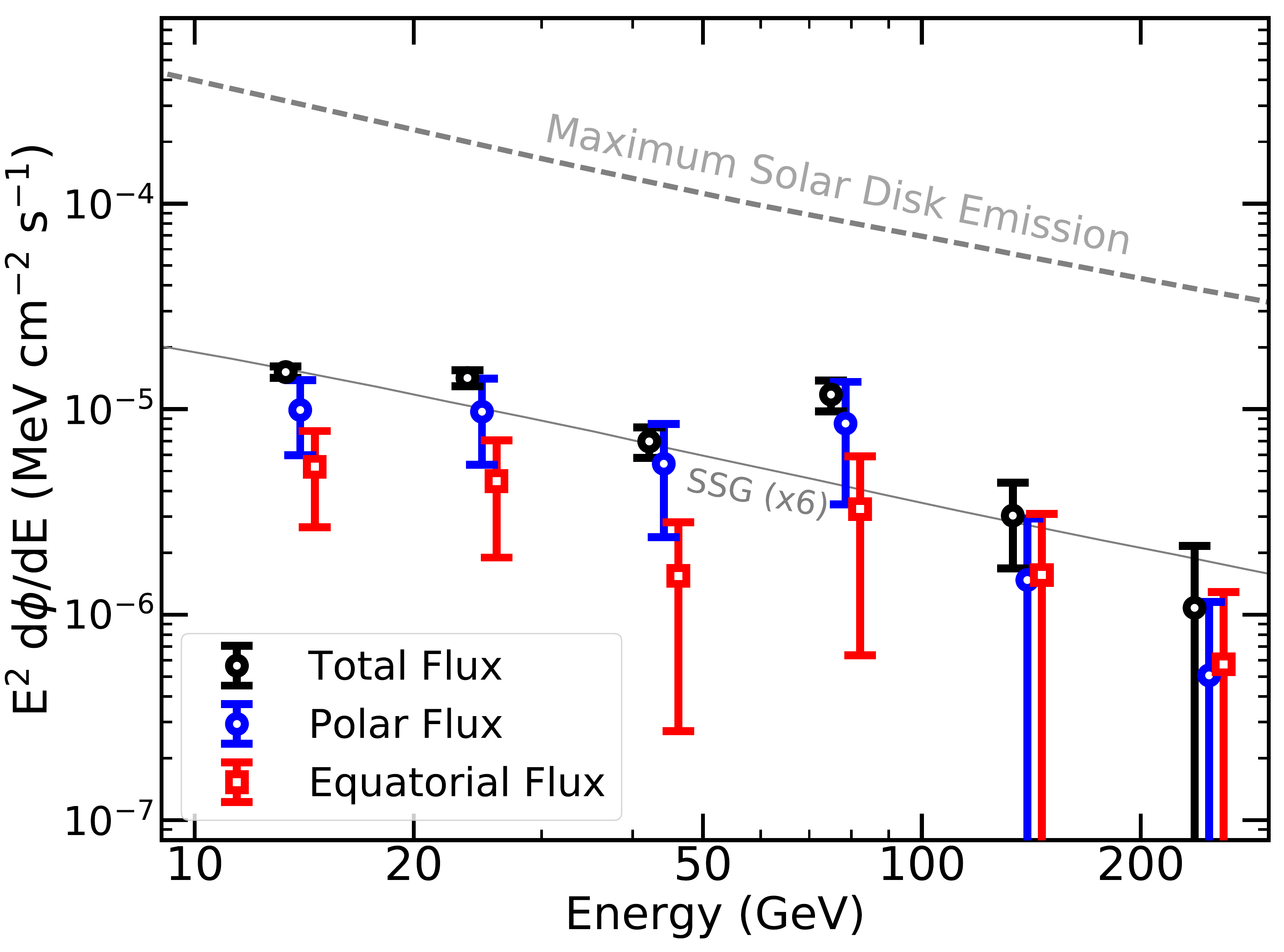}
	\caption{ Same as Figure~\ref{fig:helioprojective} (bottom), except that we do not divide the dataset into temporal periods during and after the solar minimum. For each datapoint, we utilize our standard analysis package to deconvolve the data into polar and equatorial components, to determine any independent spectral signatures of polar and equatorial emission. We find that the spectral dip appears present in both $\gamma$-rays observed near the solar equatorial plane as well as $\gamma$-rays observed near the Sun's polar regions.}
	\label{app:fig:polarequatorialspectrum}
\end{figure}

In this paper, along with our work in Ref.~\citep{paper}, we have identified both the presence of a significant spectral dip between the energies of 30--50~GeV, as well as a significant morphological shift from polar to equatorial $\gamma$-ray emission after solar minimum. It is natural to wonder whether these two feature are linked, in particular whether the spectral dip is found only for photons interacting in specific solar regions. 

We note that previous analysis should provide insight into this question. In Figure~\ref{fig:spectrumratio}, we split the $\gamma$-ray emission into two temporal bins during and after the solar minimum, finding that the spectral dip was present in events both during and after the solar minimum. Moreover, in Figure~\ref{fig:helioprojective} (bottom), we showed that the spectral dip to occur in both polar and equatorial emission components. These findings both suggest that the spectral dip is unconnected to the polar or equatorial nature of the solar emission. Unfortunately, in both cases the photon statistics are relatively poor (primarily due to the very low photon count \emph{within} the dip), thus we attempt to analyze a larger sample to test this possibility.

In Figure~\ref{app:fig:polarequatorialspectrum}, we attempt this measurement again, but combine all $\gamma$-rays into a single temporal bin to increase the event statistics. We find that, in this larger event sample, the 30--50~GeV spectral dip is still clearly identified in events from both polar and equatorial regions. We note that the dominance of polar emission over the full energy-range stems from the much longer period (7.5 years) of observations outside of solar minimum, compared to the 1.5 years of solar minimum activity. These observations indicate that the 30--50~GeV spectral dip may be a new phenomenon that is unconnected to the solar minimum cycle itself.

\bibliography{solarminimum}

\begin{thebibliography}{48}%
\makeatletter
\providecommand \@ifxundefined [1]{%
 \@ifx{#1\undefined}
}%
\providecommand \@ifnum [1]{%
 \ifnum #1\expandafter \@firstoftwo
 \else \expandafter \@secondoftwo
 \fi
}%
\providecommand \@ifx [1]{%
 \ifx #1\expandafter \@firstoftwo
 \else \expandafter \@secondoftwo
 \fi
}%
\providecommand \natexlab [1]{#1}%
\providecommand \enquote  [1]{``#1''}%
\providecommand \bibnamefont  [1]{#1}%
\providecommand \bibfnamefont [1]{#1}%
\providecommand \citenamefont [1]{#1}%
\providecommand \href@noop [0]{\@secondoftwo}%
\providecommand \href [0]{\begingroup \@sanitize@url \@href}%
\providecommand \@href[1]{\@@startlink{#1}\@@href}%
\providecommand \@@href[1]{\endgroup#1\@@endlink}%
\providecommand \@sanitize@url [0]{\catcode `\\12\catcode `\$12\catcode
  `\&12\catcode `\#12\catcode `\^12\catcode `\_12\catcode `\%12\relax}%
\providecommand \@@startlink[1]{}%
\providecommand \@@endlink[0]{}%
\providecommand \url  [0]{\begingroup\@sanitize@url \@url }%
\providecommand \@url [1]{\endgroup\@href {#1}{\urlprefix }}%
\providecommand \urlprefix  [0]{URL }%
\providecommand \Eprint [0]{\href }%
\providecommand \doibase [0]{http://dx.doi.org/}%
\providecommand \selectlanguage [0]{\@gobble}%
\providecommand \bibinfo  [0]{\@secondoftwo}%
\providecommand \bibfield  [0]{\@secondoftwo}%
\providecommand \translation [1]{[#1]}%
\providecommand \BibitemOpen [0]{}%
\providecommand \bibitemStop [0]{}%
\providecommand \bibitemNoStop [0]{.\EOS\space}%
\providecommand \EOS [0]{\spacefactor3000\relax}%
\providecommand \BibitemShut  [1]{\csname bibitem#1\endcsname}%
\let\auto@bib@innerbib\@empty
\bibitem [{\citenamefont {{Murphy}}\ \emph {et~al.}(1987)\citenamefont
  {{Murphy}}, \citenamefont {{Dermer}},\ and\ \citenamefont
  {{Ramaty}}}]{1987ApJS...63..721M}%
  \BibitemOpen
  \bibfield  {author} {\bibinfo {author} {\bibfnamefont {R.~J.}\ \bibnamefont
  {{Murphy}}}, \bibinfo {author} {\bibfnamefont {C.~D.}\ \bibnamefont
  {{Dermer}}}, \ and\ \bibinfo {author} {\bibfnamefont {R.}~\bibnamefont
  {{Ramaty}}},\ }\href {\doibase 10.1086/191180} {\bibfield  {journal}
  {\bibinfo  {journal} {\apjs}\ }\textbf {\bibinfo {volume} {63}},\ \bibinfo
  {pages} {721} (\bibinfo {year} {1987})}\BibitemShut {NoStop}%
\bibitem [{\citenamefont {{Kanbach}}\ \emph {et~al.}(1988)\citenamefont
  {{Kanbach}}, \citenamefont {{Bertsch}}, \citenamefont {{Fichtel}},
  \citenamefont {{Hartman}}, \citenamefont {{Hunter}}, \citenamefont
  {{Kniffen}}, \citenamefont {{Hughlock}}, \citenamefont {{Favale}},
  \citenamefont {{Hofstadter}},\ and\ \citenamefont
  {{Hughes}}}]{1988SSRv...49...69K}%
  \BibitemOpen
  \bibfield  {author} {\bibinfo {author} {\bibfnamefont {G.}~\bibnamefont
  {{Kanbach}}}, \bibinfo {author} {\bibfnamefont {D.~L.}\ \bibnamefont
  {{Bertsch}}}, \bibinfo {author} {\bibfnamefont {C.~E.}\ \bibnamefont
  {{Fichtel}}}, \bibinfo {author} {\bibfnamefont {R.~C.}\ \bibnamefont
  {{Hartman}}}, \bibinfo {author} {\bibfnamefont {S.~D.}\ \bibnamefont
  {{Hunter}}}, \bibinfo {author} {\bibfnamefont {D.~A.}\ \bibnamefont
  {{Kniffen}}}, \bibinfo {author} {\bibfnamefont {B.~W.}\ \bibnamefont
  {{Hughlock}}}, \bibinfo {author} {\bibfnamefont {A.}~\bibnamefont
  {{Favale}}}, \bibinfo {author} {\bibfnamefont {R.}~\bibnamefont
  {{Hofstadter}}}, \ and\ \bibinfo {author} {\bibfnamefont {E.~B.}\
  \bibnamefont {{Hughes}}},\ }\href@noop {} {\bibfield  {journal} {\bibinfo
  {journal} {\ssr}\ }\textbf {\bibinfo {volume} {49}},\ \bibinfo {pages} {69}
  (\bibinfo {year} {1988})}\BibitemShut {NoStop}%
\bibitem [{\citenamefont {{Kanbach}}\ \emph {et~al.}(1993)\citenamefont
  {{Kanbach}}, \citenamefont {{Bertsch}}, \citenamefont {{Fichtel}},
  \citenamefont {{Hartman}}, \citenamefont {{Hunter}}, \citenamefont
  {{Kniffen}}, \citenamefont {{Kwok}}, \citenamefont {{Lin}}, \citenamefont
  {{Mattox}},\ and\ \citenamefont
  {{Mayer-Hasselwander}}}]{1993A&AS...97..349K}%
  \BibitemOpen
  \bibfield  {author} {\bibinfo {author} {\bibfnamefont {G.}~\bibnamefont
  {{Kanbach}}}, \bibinfo {author} {\bibfnamefont {D.~L.}\ \bibnamefont
  {{Bertsch}}}, \bibinfo {author} {\bibfnamefont {C.~E.}\ \bibnamefont
  {{Fichtel}}}, \bibinfo {author} {\bibfnamefont {R.~C.}\ \bibnamefont
  {{Hartman}}}, \bibinfo {author} {\bibfnamefont {S.~D.}\ \bibnamefont
  {{Hunter}}}, \bibinfo {author} {\bibfnamefont {D.~A.}\ \bibnamefont
  {{Kniffen}}}, \bibinfo {author} {\bibfnamefont {P.~W.}\ \bibnamefont
  {{Kwok}}}, \bibinfo {author} {\bibfnamefont {Y.~C.}\ \bibnamefont {{Lin}}},
  \bibinfo {author} {\bibfnamefont {J.~R.}\ \bibnamefont {{Mattox}}}, \ and\
  \bibinfo {author} {\bibfnamefont {H.~A.}\ \bibnamefont
  {{Mayer-Hasselwander}}},\ }\href@noop {} {\bibfield  {journal} {\bibinfo
  {journal} {\aaps}\ }\textbf {\bibinfo {volume} {97}},\ \bibinfo {pages} {349}
  (\bibinfo {year} {1993})}\BibitemShut {NoStop}%
\bibitem [{\citenamefont {Ackermann}\ \emph {et~al.}(2014)\citenamefont
  {Ackermann} \emph {et~al.}}]{Ackermann:2013tya}%
  \BibitemOpen
  \bibfield  {author} {\bibinfo {author} {\bibfnamefont {M.}~\bibnamefont
  {Ackermann}} \emph {et~al.} (\bibinfo {collaboration} {Fermi-LAT}),\ }\href
  {\doibase 10.1088/0004-637X/787/1/15} {\bibfield  {journal} {\bibinfo
  {journal} {Astrophys. J.}\ }\textbf {\bibinfo {volume} {787}},\ \bibinfo
  {pages} {15} (\bibinfo {year} {2014})},\ \Eprint
  {http://arxiv.org/abs/1304.3749} {arXiv:1304.3749 [astro-ph.HE]} \BibitemShut
  {NoStop}%
\bibitem [{\citenamefont {Pesce-Rollins}\ \emph {et~al.}(2015)\citenamefont
  {Pesce-Rollins}, \citenamefont {Omodei}, \citenamefont {Petrosian},
  \citenamefont {Liu}, \citenamefont {da~Costa}, \citenamefont {Allafort},\
  and\ \citenamefont {Chen}}]{Pesce-Rollins:2015hpa}%
  \BibitemOpen
  \bibfield  {author} {\bibinfo {author} {\bibfnamefont {M.}~\bibnamefont
  {Pesce-Rollins}}, \bibinfo {author} {\bibfnamefont {N.}~\bibnamefont
  {Omodei}}, \bibinfo {author} {\bibfnamefont {V.}~\bibnamefont {Petrosian}},
  \bibinfo {author} {\bibfnamefont {W.}~\bibnamefont {Liu}}, \bibinfo {author}
  {\bibfnamefont {F.~R.}\ \bibnamefont {da~Costa}}, \bibinfo {author}
  {\bibfnamefont {A.}~\bibnamefont {Allafort}}, \ and\ \bibinfo {author}
  {\bibfnamefont {Q.}~\bibnamefont {Chen}},\ }\href {\doibase
  10.1088/2041-8205/805/2/L15} {\bibfield  {journal} {\bibinfo  {journal}
  {Astrophys. J.}\ }\textbf {\bibinfo {volume} {805}},\ \bibinfo {pages} {L15}
  (\bibinfo {year} {2015})},\ \Eprint {http://arxiv.org/abs/1505.03480}
  {arXiv:1505.03480 [astro-ph.SR]} \BibitemShut {NoStop}%
\bibitem [{\citenamefont {Ackermann}\ \emph {et~al.}(2017)\citenamefont
  {Ackermann} \emph {et~al.}}]{Ackermann:2017uer}%
  \BibitemOpen
  \bibfield  {author} {\bibinfo {author} {\bibfnamefont {M.}~\bibnamefont
  {Ackermann}} \emph {et~al.} (\bibinfo {collaboration} {Fermi-LAT}),\ }\href
  {\doibase 10.3847/1538-4357/835/2/219} {\bibfield  {journal} {\bibinfo
  {journal} {Astrophys. J.}\ }\textbf {\bibinfo {volume} {835}},\ \bibinfo
  {pages} {219} (\bibinfo {year} {2017})},\ \Eprint
  {http://arxiv.org/abs/1702.00577} {arXiv:1702.00577 [astro-ph.SR]}
  \BibitemShut {NoStop}%
\bibitem [{\citenamefont {Share}\ \emph {et~al.}(2017)\citenamefont {Share},
  \citenamefont {Murphy}, \citenamefont {Tolbert}, \citenamefont {Dennis},
  \citenamefont {White}, \citenamefont {Schwartz},\ and\ \citenamefont
  {Tylka}}]{Share:2017tgw}%
  \BibitemOpen
  \bibfield  {author} {\bibinfo {author} {\bibfnamefont {G.~H.}\ \bibnamefont
  {Share}}, \bibinfo {author} {\bibfnamefont {R.~J.}\ \bibnamefont {Murphy}},
  \bibinfo {author} {\bibfnamefont {A.~K.}\ \bibnamefont {Tolbert}}, \bibinfo
  {author} {\bibfnamefont {B.~R.}\ \bibnamefont {Dennis}}, \bibinfo {author}
  {\bibfnamefont {S.~M.}\ \bibnamefont {White}}, \bibinfo {author}
  {\bibfnamefont {R.~A.}\ \bibnamefont {Schwartz}}, \ and\ \bibinfo {author}
  {\bibfnamefont {A.~J.}\ \bibnamefont {Tylka}},\ }\href@noop {} {\  (\bibinfo
  {year} {2017})},\ \Eprint {http://arxiv.org/abs/1711.01511} {arXiv:1711.01511
  [astro-ph.SR]} \BibitemShut {NoStop}%
\bibitem [{\citenamefont {Ajello}\ \emph {et~al.}(2014)\citenamefont {Ajello}
  \emph {et~al.}}]{Fermi-LAT:2013cla}%
  \BibitemOpen
  \bibfield  {author} {\bibinfo {author} {\bibfnamefont {M.}~\bibnamefont
  {Ajello}} \emph {et~al.} (\bibinfo {collaboration} {Fermi-LAT}),\ }\href
  {\doibase 10.1088/0004-637X/789/1/20} {\bibfield  {journal} {\bibinfo
  {journal} {Astrophys. J.}\ }\textbf {\bibinfo {volume} {789}},\ \bibinfo
  {pages} {20} (\bibinfo {year} {2014})},\ \Eprint
  {http://arxiv.org/abs/1304.5559} {arXiv:1304.5559 [astro-ph.HE]} \BibitemShut
  {NoStop}%
\bibitem [{\citenamefont {{Abdo}}\ \emph {et~al.}(2011)\citenamefont {{Abdo}},
  \citenamefont {{Ackermann}}, \citenamefont {{Ajello}}, \citenamefont
  {{Baldini}}, \citenamefont {{Ballet}}, \citenamefont {{Barbiellini}},
  \citenamefont {{Bastieri}}, \citenamefont {{Bechtol}},\ and\ \citenamefont
  {{et al.}}}]{2011ApJ...734..116A}%
  \BibitemOpen
  \bibfield  {author} {\bibinfo {author} {\bibfnamefont {A.~A.}\ \bibnamefont
  {{Abdo}}}, \bibinfo {author} {\bibfnamefont {M.}~\bibnamefont {{Ackermann}}},
  \bibinfo {author} {\bibfnamefont {M.}~\bibnamefont {{Ajello}}}, \bibinfo
  {author} {\bibfnamefont {L.}~\bibnamefont {{Baldini}}}, \bibinfo {author}
  {\bibfnamefont {J.}~\bibnamefont {{Ballet}}}, \bibinfo {author}
  {\bibfnamefont {G.}~\bibnamefont {{Barbiellini}}}, \bibinfo {author}
  {\bibfnamefont {D.}~\bibnamefont {{Bastieri}}}, \bibinfo {author}
  {\bibfnamefont {K.}~\bibnamefont {{Bechtol}}}, \ and\ \bibinfo {author}
  {\bibnamefont {{et al.}}},\ }\href {\doibase 10.1088/0004-637X/734/2/116}
  {\bibfield  {journal} {\bibinfo  {journal} {\apj}\ }\textbf {\bibinfo
  {volume} {734}},\ \bibinfo {eid} {116} (\bibinfo {year} {2011})},\ \Eprint
  {http://arxiv.org/abs/1104.2093} {arXiv:1104.2093 [astro-ph.HE]} \BibitemShut
  {NoStop}%
\bibitem [{\citenamefont {{Moskalenko}}\ \emph {et~al.}(2006)\citenamefont
  {{Moskalenko}}, \citenamefont {{Porter}},\ and\ \citenamefont
  {{Digel}}}]{2006ApJ...652L..65M}%
  \BibitemOpen
  \bibfield  {author} {\bibinfo {author} {\bibfnamefont {I.~V.}\ \bibnamefont
  {{Moskalenko}}}, \bibinfo {author} {\bibfnamefont {T.~A.}\ \bibnamefont
  {{Porter}}}, \ and\ \bibinfo {author} {\bibfnamefont {S.~W.}\ \bibnamefont
  {{Digel}}},\ }\href {\doibase 10.1086/509916} {\bibfield  {journal} {\bibinfo
   {journal} {\apjl}\ }\textbf {\bibinfo {volume} {652}},\ \bibinfo {pages}
  {L65} (\bibinfo {year} {2006})},\ \Eprint
  {http://arxiv.org/abs/astro-ph/0607521} {astro-ph/0607521} \BibitemShut
  {NoStop}%
\bibitem [{\citenamefont {{Orlando}}\ and\ \citenamefont
  {{Strong}}(2007)}]{2007Ap&SS.309..359O}%
  \BibitemOpen
  \bibfield  {author} {\bibinfo {author} {\bibfnamefont {E.}~\bibnamefont
  {{Orlando}}}\ and\ \bibinfo {author} {\bibfnamefont {A.~W.}\ \bibnamefont
  {{Strong}}},\ }\href {\doibase 10.1007/s10509-007-9457-0} {\bibfield
  {journal} {\bibinfo  {journal} {\apss}\ }\textbf {\bibinfo {volume} {309}},\
  \bibinfo {pages} {359} (\bibinfo {year} {2007})},\ \Eprint
  {http://arxiv.org/abs/astro-ph/0607563} {astro-ph/0607563} \BibitemShut
  {NoStop}%
\bibitem [{\citenamefont {Orlando}\ \emph {et~al.}(2017)\citenamefont
  {Orlando}, \citenamefont {Giglietto}, \citenamefont {Moskalenko},
  \citenamefont {Raino'},\ and\ \citenamefont {Strong}}]{Orlando:2017iyc}%
  \BibitemOpen
  \bibfield  {author} {\bibinfo {author} {\bibfnamefont {E.}~\bibnamefont
  {Orlando}}, \bibinfo {author} {\bibfnamefont {N.}~\bibnamefont {Giglietto}},
  \bibinfo {author} {\bibfnamefont {I.}~\bibnamefont {Moskalenko}}, \bibinfo
  {author} {\bibfnamefont {S.}~\bibnamefont {Raino'}}, \ and\ \bibinfo {author}
  {\bibfnamefont {A.}~\bibnamefont {Strong}},\ }\bibfield  {booktitle} {\emph
  {\bibinfo {booktitle} {{Proceedings, 35th International Cosmic Ray Conference
  (ICRC 2017): Bexco, Busan, Korea, July 12-20, 2017}}},\ }\href@noop {}
  {\bibfield  {journal} {\bibinfo  {journal} {PoS}\ }\textbf {\bibinfo {volume}
  {ICRC2017}},\ \bibinfo {pages} {693} (\bibinfo {year} {2017})},\ \Eprint
  {http://arxiv.org/abs/1712.09745} {arXiv:1712.09745 [astro-ph.HE]}
  \BibitemShut {NoStop}%
\bibitem [{\citenamefont {Ng}\ \emph {et~al.}(2016)\citenamefont {Ng},
  \citenamefont {Beacom}, \citenamefont {Peter},\ and\ \citenamefont
  {Rott}}]{Ng:2015gya}%
  \BibitemOpen
  \bibfield  {author} {\bibinfo {author} {\bibfnamefont {K.~C.~Y.}\
  \bibnamefont {Ng}}, \bibinfo {author} {\bibfnamefont {J.~F.}\ \bibnamefont
  {Beacom}}, \bibinfo {author} {\bibfnamefont {A.~H.~G.}\ \bibnamefont
  {Peter}}, \ and\ \bibinfo {author} {\bibfnamefont {C.}~\bibnamefont {Rott}},\
  }\href {\doibase 10.1103/PhysRevD.94.023004} {\bibfield  {journal} {\bibinfo
  {journal} {Phys. Rev.}\ }\textbf {\bibinfo {volume} {D94}},\ \bibinfo {pages}
  {023004} (\bibinfo {year} {2016})},\ \Eprint
  {http://arxiv.org/abs/1508.06276} {arXiv:1508.06276 [astro-ph.HE]}
  \BibitemShut {NoStop}%
\bibitem [{\citenamefont {Orlando}\ and\ \citenamefont
  {Strong}(2008)}]{Orlando:2008uk}%
  \BibitemOpen
  \bibfield  {author} {\bibinfo {author} {\bibfnamefont {E.}~\bibnamefont
  {Orlando}}\ and\ \bibinfo {author} {\bibfnamefont {A.~W.}\ \bibnamefont
  {Strong}},\ }\href {\doibase 10.1051/0004-6361:20078817} {\bibfield
  {journal} {\bibinfo  {journal} {Astron. Astrophys.}\ }\textbf {\bibinfo
  {volume} {480}},\ \bibinfo {pages} {847} (\bibinfo {year} {2008})},\ \Eprint
  {http://arxiv.org/abs/0801.2178} {arXiv:0801.2178 [astro-ph]} \BibitemShut
  {NoStop}%
\bibitem [{\citenamefont {Zhou}\ \emph {et~al.}(2017)\citenamefont {Zhou},
  \citenamefont {Ng}, \citenamefont {Beacom},\ and\ \citenamefont
  {Peter}}]{Zhou:2016ljf}%
  \BibitemOpen
  \bibfield  {author} {\bibinfo {author} {\bibfnamefont {B.}~\bibnamefont
  {Zhou}}, \bibinfo {author} {\bibfnamefont {K.~C.~Y.}\ \bibnamefont {Ng}},
  \bibinfo {author} {\bibfnamefont {J.~F.}\ \bibnamefont {Beacom}}, \ and\
  \bibinfo {author} {\bibfnamefont {A.~H.~G.}\ \bibnamefont {Peter}},\ }\href
  {\doibase 10.1103/PhysRevD.96.023015} {\bibfield  {journal} {\bibinfo
  {journal} {Phys. Rev.}\ }\textbf {\bibinfo {volume} {D96}},\ \bibinfo {pages}
  {023015} (\bibinfo {year} {2017})},\ \Eprint
  {http://arxiv.org/abs/1612.02420} {arXiv:1612.02420 [astro-ph.HE]}
  \BibitemShut {NoStop}%
\bibitem [{\citenamefont {Seckel}\ \emph {et~al.}(1991)\citenamefont {Seckel},
  \citenamefont {Stanev},\ and\ \citenamefont {Gaisser}}]{Seckel:1991ffa}%
  \BibitemOpen
  \bibfield  {author} {\bibinfo {author} {\bibfnamefont {D.}~\bibnamefont
  {Seckel}}, \bibinfo {author} {\bibfnamefont {T.}~\bibnamefont {Stanev}}, \
  and\ \bibinfo {author} {\bibfnamefont {T.~K.}\ \bibnamefont {Gaisser}},\
  }\href {\doibase 10.1086/170753} {\bibfield  {journal} {\bibinfo  {journal}
  {Astrophys. J.}\ }\textbf {\bibinfo {volume} {382}},\ \bibinfo {pages} {652}
  (\bibinfo {year} {1991})}\BibitemShut {NoStop}%
\bibitem [{\citenamefont {Tang}\ \emph {et~al.}(2018)\citenamefont {Tang},
  \citenamefont {Ng}, \citenamefont {Linden}, \citenamefont {Zhou},
  \citenamefont {Beacom},\ and\ \citenamefont {Peter}}]{paper}%
  \BibitemOpen
  \bibfield  {author} {\bibinfo {author} {\bibfnamefont {Q.-W.}\ \bibnamefont
  {Tang}}, \bibinfo {author} {\bibfnamefont {K.~C.~Y.}\ \bibnamefont {Ng}},
  \bibinfo {author} {\bibfnamefont {T.}~\bibnamefont {Linden}}, \bibinfo
  {author} {\bibfnamefont {B.}~\bibnamefont {Zhou}}, \bibinfo {author}
  {\bibfnamefont {J.~F.}\ \bibnamefont {Beacom}}, \ and\ \bibinfo {author}
  {\bibfnamefont {A.~H.~G.}\ \bibnamefont {Peter}},\ }\href@noop {} {\
  (\bibinfo {year} {2018})},\ \Eprint {http://arxiv.org/abs/1801.XXXXX}
  {arXiv:1801.XXXXX} \BibitemShut {NoStop}%
\bibitem [{\citenamefont {Atwood}\ \emph {et~al.}(2009)\citenamefont {Atwood}
  \emph {et~al.}}]{Atwood:2009ez}%
  \BibitemOpen
  \bibfield  {author} {\bibinfo {author} {\bibfnamefont {W.~B.}\ \bibnamefont
  {Atwood}} \emph {et~al.} (\bibinfo {collaboration} {Fermi-LAT}),\ }\href
  {\doibase 10.1088/0004-637X/697/2/1071} {\bibfield  {journal} {\bibinfo
  {journal} {Astrophys. J.}\ }\textbf {\bibinfo {volume} {697}},\ \bibinfo
  {pages} {1071} (\bibinfo {year} {2009})},\ \Eprint
  {http://arxiv.org/abs/0902.1089} {arXiv:0902.1089 [astro-ph.IM]} \BibitemShut
  {NoStop}%
\bibitem [{\citenamefont {Abeysekara}\ \emph {et~al.}(2013)\citenamefont
  {Abeysekara} \emph {et~al.}}]{Abeysekara:2013tza}%
  \BibitemOpen
  \bibfield  {author} {\bibinfo {author} {\bibfnamefont {A.~U.}\ \bibnamefont
  {Abeysekara}} \emph {et~al.},\ }\href {\doibase
  10.1016/j.astropartphys.2013.08.002} {\bibfield  {journal} {\bibinfo
  {journal} {Astropart. Phys.}\ }\textbf {\bibinfo {volume} {50-52}},\ \bibinfo
  {pages} {26} (\bibinfo {year} {2013})},\ \Eprint
  {http://arxiv.org/abs/1306.5800} {arXiv:1306.5800 [astro-ph.HE]} \BibitemShut
  {NoStop}%
\bibitem [{\citenamefont {Aartsen}\ \emph {et~al.}(2013)\citenamefont {Aartsen}
  \emph {et~al.}}]{Aartsen:2012kia}%
  \BibitemOpen
  \bibfield  {author} {\bibinfo {author} {\bibfnamefont {M.~G.}\ \bibnamefont
  {Aartsen}} \emph {et~al.} (\bibinfo {collaboration} {IceCube}),\ }\href
  {\doibase 10.1103/PhysRevLett.110.131302} {\bibfield  {journal} {\bibinfo
  {journal} {Phys. Rev. Lett.}\ }\textbf {\bibinfo {volume} {110}},\ \bibinfo
  {pages} {131302} (\bibinfo {year} {2013})},\ \Eprint
  {http://arxiv.org/abs/1212.4097} {arXiv:1212.4097 [astro-ph.HE]} \BibitemShut
  {NoStop}%
\bibitem [{\citenamefont {Batell}\ \emph {et~al.}(2010)\citenamefont {Batell},
  \citenamefont {Pospelov}, \citenamefont {Ritz},\ and\ \citenamefont
  {Shang}}]{Batell:2009zp}%
  \BibitemOpen
  \bibfield  {author} {\bibinfo {author} {\bibfnamefont {B.}~\bibnamefont
  {Batell}}, \bibinfo {author} {\bibfnamefont {M.}~\bibnamefont {Pospelov}},
  \bibinfo {author} {\bibfnamefont {A.}~\bibnamefont {Ritz}}, \ and\ \bibinfo
  {author} {\bibfnamefont {Y.}~\bibnamefont {Shang}},\ }\href {\doibase
  10.1103/PhysRevD.81.075004} {\bibfield  {journal} {\bibinfo  {journal} {Phys.
  Rev.}\ }\textbf {\bibinfo {volume} {D81}},\ \bibinfo {pages} {075004}
  (\bibinfo {year} {2010})},\ \Eprint {http://arxiv.org/abs/0910.1567}
  {arXiv:0910.1567 [hep-ph]} \BibitemShut {NoStop}%
\bibitem [{\citenamefont {Schuster}\ \emph {et~al.}(2010)\citenamefont
  {Schuster}, \citenamefont {Toro},\ and\ \citenamefont
  {Yavin}}]{Schuster:2009au}%
  \BibitemOpen
  \bibfield  {author} {\bibinfo {author} {\bibfnamefont {P.}~\bibnamefont
  {Schuster}}, \bibinfo {author} {\bibfnamefont {N.}~\bibnamefont {Toro}}, \
  and\ \bibinfo {author} {\bibfnamefont {I.}~\bibnamefont {Yavin}},\ }\href
  {\doibase 10.1103/PhysRevD.81.016002} {\bibfield  {journal} {\bibinfo
  {journal} {Phys. Rev.}\ }\textbf {\bibinfo {volume} {D81}},\ \bibinfo {pages}
  {016002} (\bibinfo {year} {2010})},\ \Eprint {http://arxiv.org/abs/0910.1602}
  {arXiv:0910.1602 [hep-ph]} \BibitemShut {NoStop}%
\bibitem [{\citenamefont {Meade}\ \emph {et~al.}(2010)\citenamefont {Meade},
  \citenamefont {Nussinov}, \citenamefont {Papucci},\ and\ \citenamefont
  {Volansky}}]{Meade:2009mu}%
  \BibitemOpen
  \bibfield  {author} {\bibinfo {author} {\bibfnamefont {P.}~\bibnamefont
  {Meade}}, \bibinfo {author} {\bibfnamefont {S.}~\bibnamefont {Nussinov}},
  \bibinfo {author} {\bibfnamefont {M.}~\bibnamefont {Papucci}}, \ and\
  \bibinfo {author} {\bibfnamefont {T.}~\bibnamefont {Volansky}},\ }\href
  {\doibase 10.1007/JHEP06(2010)029} {\bibfield  {journal} {\bibinfo  {journal}
  {JHEP}\ }\textbf {\bibinfo {volume} {06}},\ \bibinfo {pages} {029} (\bibinfo
  {year} {2010})},\ \Eprint {http://arxiv.org/abs/0910.4160} {arXiv:0910.4160
  [hep-ph]} \BibitemShut {NoStop}%
\bibitem [{\citenamefont {Bell}\ and\ \citenamefont
  {Petraki}(2011)}]{Bell:2011sn}%
  \BibitemOpen
  \bibfield  {author} {\bibinfo {author} {\bibfnamefont {N.~F.}\ \bibnamefont
  {Bell}}\ and\ \bibinfo {author} {\bibfnamefont {K.}~\bibnamefont {Petraki}},\
  }\href {\doibase 10.1088/1475-7516/2011/04/003} {\bibfield  {journal}
  {\bibinfo  {journal} {JCAP}\ }\textbf {\bibinfo {volume} {1104}},\ \bibinfo
  {pages} {003} (\bibinfo {year} {2011})},\ \Eprint
  {http://arxiv.org/abs/1102.2958} {arXiv:1102.2958 [hep-ph]} \BibitemShut
  {NoStop}%
\bibitem [{\citenamefont {Feng}\ \emph {et~al.}(2016)\citenamefont {Feng},
  \citenamefont {Smolinsky},\ and\ \citenamefont {Tanedo}}]{Feng:2016ijc}%
  \BibitemOpen
  \bibfield  {author} {\bibinfo {author} {\bibfnamefont {J.~L.}\ \bibnamefont
  {Feng}}, \bibinfo {author} {\bibfnamefont {J.}~\bibnamefont {Smolinsky}}, \
  and\ \bibinfo {author} {\bibfnamefont {P.}~\bibnamefont {Tanedo}},\ }\href
  {\doibase 10.1103/PhysRevD.93.115036, 10.1103/PhysRevD.96.099903} {\bibfield
  {journal} {\bibinfo  {journal} {Phys. Rev.}\ }\textbf {\bibinfo {volume}
  {D93}},\ \bibinfo {pages} {115036} (\bibinfo {year} {2016})},\ \bibinfo
  {note} {[Erratum: Phys. Rev.D96,no.9,099903(2017)]},\ \Eprint
  {http://arxiv.org/abs/1602.01465} {arXiv:1602.01465 [hep-ph]} \BibitemShut
  {NoStop}%
\bibitem [{\citenamefont {Leane}\ \emph {et~al.}(2017)\citenamefont {Leane},
  \citenamefont {Ng},\ and\ \citenamefont {Beacom}}]{Leane:2017vag}%
  \BibitemOpen
  \bibfield  {author} {\bibinfo {author} {\bibfnamefont {R.~K.}\ \bibnamefont
  {Leane}}, \bibinfo {author} {\bibfnamefont {K.~C.~Y.}\ \bibnamefont {Ng}}, \
  and\ \bibinfo {author} {\bibfnamefont {J.~F.}\ \bibnamefont {Beacom}},\
  }\href {\doibase 10.1103/PhysRevD.95.123016} {\bibfield  {journal} {\bibinfo
  {journal} {Phys. Rev.}\ }\textbf {\bibinfo {volume} {D95}},\ \bibinfo {pages}
  {123016} (\bibinfo {year} {2017})},\ \Eprint
  {http://arxiv.org/abs/1703.04629} {arXiv:1703.04629 [astro-ph.HE]}
  \BibitemShut {NoStop}%
\bibitem [{\citenamefont {Argüelles}\ \emph {et~al.}(2017)\citenamefont
  {Argüelles}, \citenamefont {de~Wasseige}, \citenamefont {Fedynitch},\ and\
  \citenamefont {Jones}}]{Arguelles:2017eao}%
  \BibitemOpen
  \bibfield  {author} {\bibinfo {author} {\bibfnamefont {C.~A.}\ \bibnamefont
  {Argüelles}}, \bibinfo {author} {\bibfnamefont {G.}~\bibnamefont
  {de~Wasseige}}, \bibinfo {author} {\bibfnamefont {A.}~\bibnamefont
  {Fedynitch}}, \ and\ \bibinfo {author} {\bibfnamefont {B.~J.~P.}\
  \bibnamefont {Jones}},\ }\href {\doibase 10.1088/1475-7516/2017/07/024}
  {\bibfield  {journal} {\bibinfo  {journal} {JCAP}\ }\textbf {\bibinfo
  {volume} {1707}},\ \bibinfo {pages} {024} (\bibinfo {year} {2017})},\ \Eprint
  {http://arxiv.org/abs/1703.07798} {arXiv:1703.07798 [astro-ph.HE]}
  \BibitemShut {NoStop}%
\bibitem [{\citenamefont {Edsjo}\ \emph {et~al.}(2017)\citenamefont {Edsjo},
  \citenamefont {Elevant}, \citenamefont {Enberg},\ and\ \citenamefont
  {Niblaeus}}]{Edsjo:2017kjk}%
  \BibitemOpen
  \bibfield  {author} {\bibinfo {author} {\bibfnamefont {J.}~\bibnamefont
  {Edsjo}}, \bibinfo {author} {\bibfnamefont {J.}~\bibnamefont {Elevant}},
  \bibinfo {author} {\bibfnamefont {R.}~\bibnamefont {Enberg}}, \ and\ \bibinfo
  {author} {\bibfnamefont {C.}~\bibnamefont {Niblaeus}},\ }\href {\doibase
  10.1088/1475-7516/2017/06/033} {\bibfield  {journal} {\bibinfo  {journal}
  {JCAP}\ }\textbf {\bibinfo {volume} {1706}},\ \bibinfo {pages} {033}
  (\bibinfo {year} {2017})},\ \Eprint {http://arxiv.org/abs/1704.02892}
  {arXiv:1704.02892 [astro-ph.HE]} \BibitemShut {NoStop}%
\bibitem [{\citenamefont {Ng}\ \emph {et~al.}(2017)\citenamefont {Ng},
  \citenamefont {Beacom}, \citenamefont {Peter},\ and\ \citenamefont
  {Rott}}]{Ng:2017aur}%
  \BibitemOpen
  \bibfield  {author} {\bibinfo {author} {\bibfnamefont {K.~C.~Y.}\
  \bibnamefont {Ng}}, \bibinfo {author} {\bibfnamefont {J.~F.}\ \bibnamefont
  {Beacom}}, \bibinfo {author} {\bibfnamefont {A.~H.~G.}\ \bibnamefont
  {Peter}}, \ and\ \bibinfo {author} {\bibfnamefont {C.}~\bibnamefont {Rott}},\
  }\href {\doibase 10.1103/PhysRevD.96.103006} {\bibfield  {journal} {\bibinfo
  {journal} {Phys. Rev.}\ }\textbf {\bibinfo {volume} {D96}},\ \bibinfo {pages}
  {103006} (\bibinfo {year} {2017})},\ \Eprint
  {http://arxiv.org/abs/1703.10280} {arXiv:1703.10280 [astro-ph.HE]}
  \BibitemShut {NoStop}%
\bibitem [{\citenamefont {{SunPy Community}}\ \emph {et~al.}(2015)\citenamefont
  {{SunPy Community}}, \citenamefont {{Mumford}}, \citenamefont {{Christe}},
  \citenamefont {{P{\'e}rez-Su{\'a}rez}}, \citenamefont {{Ireland}},
  \citenamefont {{Shih}}, \citenamefont {{Inglis}}, \citenamefont {{Liedtke}},
  \citenamefont {{Hewett}}, \citenamefont {{Mayer}}, \citenamefont {{Hughitt}},
  \citenamefont {{Freij}}, \citenamefont {{Meszaros}}, \citenamefont
  {{Bennett}}, \citenamefont {{Malocha}}, \citenamefont {{Evans}},
  \citenamefont {{Agrawal}}, \citenamefont {{Leonard}}, \citenamefont
  {{Robitaille}}, \citenamefont {{Mampaey}}, \citenamefont {{Campos-Rozo}},\
  and\ \citenamefont {{Kirk}}}]{2015CS&D....8a4009S}%
  \BibitemOpen
  \bibfield  {author} {\bibinfo {author} {\bibnamefont {{SunPy Community}}},
  \bibinfo {author} {\bibfnamefont {S.~J.}\ \bibnamefont {{Mumford}}}, \bibinfo
  {author} {\bibfnamefont {S.}~\bibnamefont {{Christe}}}, \bibinfo {author}
  {\bibfnamefont {D.}~\bibnamefont {{P{\'e}rez-Su{\'a}rez}}}, \bibinfo {author}
  {\bibfnamefont {J.}~\bibnamefont {{Ireland}}}, \bibinfo {author}
  {\bibfnamefont {A.~Y.}\ \bibnamefont {{Shih}}}, \bibinfo {author}
  {\bibfnamefont {A.~R.}\ \bibnamefont {{Inglis}}}, \bibinfo {author}
  {\bibfnamefont {S.}~\bibnamefont {{Liedtke}}}, \bibinfo {author}
  {\bibfnamefont {R.~J.}\ \bibnamefont {{Hewett}}}, \bibinfo {author}
  {\bibfnamefont {F.}~\bibnamefont {{Mayer}}}, \bibinfo {author} {\bibfnamefont
  {K.}~\bibnamefont {{Hughitt}}}, \bibinfo {author} {\bibfnamefont
  {N.}~\bibnamefont {{Freij}}}, \bibinfo {author} {\bibfnamefont
  {T.}~\bibnamefont {{Meszaros}}}, \bibinfo {author} {\bibfnamefont {S.~M.}\
  \bibnamefont {{Bennett}}}, \bibinfo {author} {\bibfnamefont {M.}~\bibnamefont
  {{Malocha}}}, \bibinfo {author} {\bibfnamefont {J.}~\bibnamefont {{Evans}}},
  \bibinfo {author} {\bibfnamefont {A.}~\bibnamefont {{Agrawal}}}, \bibinfo
  {author} {\bibfnamefont {A.~J.}\ \bibnamefont {{Leonard}}}, \bibinfo {author}
  {\bibfnamefont {T.~P.}\ \bibnamefont {{Robitaille}}}, \bibinfo {author}
  {\bibfnamefont {B.}~\bibnamefont {{Mampaey}}}, \bibinfo {author}
  {\bibfnamefont {J.~I.}\ \bibnamefont {{Campos-Rozo}}}, \ and\ \bibinfo
  {author} {\bibfnamefont {M.~S.}\ \bibnamefont {{Kirk}}},\ }\href {\doibase
  10.1088/1749-4699/8/1/014009} {\bibfield  {journal} {\bibinfo  {journal}
  {Computational Science and Discovery}\ }\textbf {\bibinfo {volume} {8}},\
  \bibinfo {eid} {014009} (\bibinfo {year} {2015})},\ \Eprint
  {http://arxiv.org/abs/1505.02563} {arXiv:1505.02563 [astro-ph.IM]}
  \BibitemShut {NoStop}%
\bibitem [{\citenamefont {{Astropy Collaboration}}\ \emph
  {et~al.}(2013)\citenamefont {{Astropy Collaboration}}, \citenamefont
  {{Robitaille}}, \citenamefont {{Tollerud}}, \citenamefont {{Greenfield}},
  \citenamefont {{Droettboom}}, \citenamefont {{Bray}}, \citenamefont
  {{Aldcroft}}, \citenamefont {{Davis}}, \citenamefont {{Ginsburg}},
  \citenamefont {{Price-Whelan}}, \citenamefont {{Kerzendorf}}, \citenamefont
  {{Conley}}, \citenamefont {{Crighton}}, \citenamefont {{Barbary}},
  \citenamefont {{Muna}}, \citenamefont {{Ferguson}}, \citenamefont
  {{Grollier}}, \citenamefont {{Parikh}}, \citenamefont {{Nair}}, \citenamefont
  {{Unther}}, \citenamefont {{Deil}}, \citenamefont {{Woillez}}, \citenamefont
  {{Conseil}}, \citenamefont {{Kramer}}, \citenamefont {{Turner}},
  \citenamefont {{Singer}}, \citenamefont {{Fox}}, \citenamefont {{Weaver}},
  \citenamefont {{Zabalza}}, \citenamefont {{Edwards}}, \citenamefont {{Azalee
  Bostroem}}, \citenamefont {{Burke}}, \citenamefont {{Casey}}, \citenamefont
  {{Crawford}}, \citenamefont {{Dencheva}}, \citenamefont {{Ely}},
  \citenamefont {{Jenness}}, \citenamefont {{Labrie}}, \citenamefont {{Lim}},
  \citenamefont {{Pierfederici}}, \citenamefont {{Pontzen}}, \citenamefont
  {{Ptak}}, \citenamefont {{Refsdal}}, \citenamefont {{Servillat}},\ and\
  \citenamefont {{Streicher}}}]{2013A&A...558A..33A}%
  \BibitemOpen
  \bibfield  {author} {\bibinfo {author} {\bibnamefont {{Astropy
  Collaboration}}}, \bibinfo {author} {\bibfnamefont {T.~P.}\ \bibnamefont
  {{Robitaille}}}, \bibinfo {author} {\bibfnamefont {E.~J.}\ \bibnamefont
  {{Tollerud}}}, \bibinfo {author} {\bibfnamefont {P.}~\bibnamefont
  {{Greenfield}}}, \bibinfo {author} {\bibfnamefont {M.}~\bibnamefont
  {{Droettboom}}}, \bibinfo {author} {\bibfnamefont {E.}~\bibnamefont
  {{Bray}}}, \bibinfo {author} {\bibfnamefont {T.}~\bibnamefont {{Aldcroft}}},
  \bibinfo {author} {\bibfnamefont {M.}~\bibnamefont {{Davis}}}, \bibinfo
  {author} {\bibfnamefont {A.}~\bibnamefont {{Ginsburg}}}, \bibinfo {author}
  {\bibfnamefont {A.~M.}\ \bibnamefont {{Price-Whelan}}}, \bibinfo {author}
  {\bibfnamefont {W.~E.}\ \bibnamefont {{Kerzendorf}}}, \bibinfo {author}
  {\bibfnamefont {A.}~\bibnamefont {{Conley}}}, \bibinfo {author}
  {\bibfnamefont {N.}~\bibnamefont {{Crighton}}}, \bibinfo {author}
  {\bibfnamefont {K.}~\bibnamefont {{Barbary}}}, \bibinfo {author}
  {\bibfnamefont {D.}~\bibnamefont {{Muna}}}, \bibinfo {author} {\bibfnamefont
  {H.}~\bibnamefont {{Ferguson}}}, \bibinfo {author} {\bibfnamefont
  {F.}~\bibnamefont {{Grollier}}}, \bibinfo {author} {\bibfnamefont {M.~M.}\
  \bibnamefont {{Parikh}}}, \bibinfo {author} {\bibfnamefont {P.~H.}\
  \bibnamefont {{Nair}}}, \bibinfo {author} {\bibfnamefont {H.~M.}\
  \bibnamefont {{Unther}}}, \bibinfo {author} {\bibfnamefont {C.}~\bibnamefont
  {{Deil}}}, \bibinfo {author} {\bibfnamefont {J.}~\bibnamefont {{Woillez}}},
  \bibinfo {author} {\bibfnamefont {S.}~\bibnamefont {{Conseil}}}, \bibinfo
  {author} {\bibfnamefont {R.}~\bibnamefont {{Kramer}}}, \bibinfo {author}
  {\bibfnamefont {J.~E.~H.}\ \bibnamefont {{Turner}}}, \bibinfo {author}
  {\bibfnamefont {L.}~\bibnamefont {{Singer}}}, \bibinfo {author}
  {\bibfnamefont {R.}~\bibnamefont {{Fox}}}, \bibinfo {author} {\bibfnamefont
  {B.~A.}\ \bibnamefont {{Weaver}}}, \bibinfo {author} {\bibfnamefont
  {V.}~\bibnamefont {{Zabalza}}}, \bibinfo {author} {\bibfnamefont {Z.~I.}\
  \bibnamefont {{Edwards}}}, \bibinfo {author} {\bibfnamefont {K.}~\bibnamefont
  {{Azalee Bostroem}}}, \bibinfo {author} {\bibfnamefont {D.~J.}\ \bibnamefont
  {{Burke}}}, \bibinfo {author} {\bibfnamefont {A.~R.}\ \bibnamefont
  {{Casey}}}, \bibinfo {author} {\bibfnamefont {S.~M.}\ \bibnamefont
  {{Crawford}}}, \bibinfo {author} {\bibfnamefont {N.}~\bibnamefont
  {{Dencheva}}}, \bibinfo {author} {\bibfnamefont {J.}~\bibnamefont {{Ely}}},
  \bibinfo {author} {\bibfnamefont {T.}~\bibnamefont {{Jenness}}}, \bibinfo
  {author} {\bibfnamefont {K.}~\bibnamefont {{Labrie}}}, \bibinfo {author}
  {\bibfnamefont {P.~L.}\ \bibnamefont {{Lim}}}, \bibinfo {author}
  {\bibfnamefont {F.}~\bibnamefont {{Pierfederici}}}, \bibinfo {author}
  {\bibfnamefont {A.}~\bibnamefont {{Pontzen}}}, \bibinfo {author}
  {\bibfnamefont {A.}~\bibnamefont {{Ptak}}}, \bibinfo {author} {\bibfnamefont
  {B.}~\bibnamefont {{Refsdal}}}, \bibinfo {author} {\bibfnamefont
  {M.}~\bibnamefont {{Servillat}}}, \ and\ \bibinfo {author} {\bibfnamefont
  {O.}~\bibnamefont {{Streicher}}},\ }\href {\doibase
  10.1051/0004-6361/201322068} {\bibfield  {journal} {\bibinfo  {journal}
  {\aap}\ }\textbf {\bibinfo {volume} {558}},\ \bibinfo {eid} {A33} (\bibinfo
  {year} {2013})},\ \Eprint {http://arxiv.org/abs/1307.6212} {arXiv:1307.6212
  [astro-ph.IM]} \BibitemShut {NoStop}%
\bibitem [{\citenamefont {Byrne}\ \emph {et~al.}(2010)\citenamefont {Byrne},
  \citenamefont {Maloney}, \citenamefont {McAteer}, \citenamefont {Refojo},\
  and\ \citenamefont {Gallagher}}]{Byrne:2010rz}%
  \BibitemOpen
  \bibfield  {author} {\bibinfo {author} {\bibfnamefont {J.~P.}\ \bibnamefont
  {Byrne}}, \bibinfo {author} {\bibfnamefont {S.~A.}\ \bibnamefont {Maloney}},
  \bibinfo {author} {\bibfnamefont {R.~T.~J.}\ \bibnamefont {McAteer}},
  \bibinfo {author} {\bibfnamefont {J.~M.}\ \bibnamefont {Refojo}}, \ and\
  \bibinfo {author} {\bibfnamefont {P.~T.}\ \bibnamefont {Gallagher}},\ }\href
  {\doibase 10.1038/ncomms1077} {\bibfield  {journal} {\bibinfo  {journal}
  {Nature Commun.}\ }\textbf {\bibinfo {volume} {1}},\ \bibinfo {pages} {74}
  (\bibinfo {year} {2010})},\ \Eprint {http://arxiv.org/abs/1010.0643}
  {arXiv:1010.0643 [astro-ph.SR]} \BibitemShut {NoStop}%
\bibitem [{\citenamefont {Byrne}(2012)}]{Byrne:2012nk}%
  \BibitemOpen
  \bibfield  {author} {\bibinfo {author} {\bibfnamefont {J.~P.}\ \bibnamefont
  {Byrne}},\ }\emph {\bibinfo {title} {{The Kinematics and Morphology of Solar
  Coronal Mass Ejections}}},\ \href
  {http://inspirehep.net/record/1089454/files/arXiv:1202.4005.pdf} {Ph.D.
  thesis} (\bibinfo {year} {2012}),\ \Eprint {http://arxiv.org/abs/1202.4005}
  {arXiv:1202.4005 [astro-ph.SR]} \BibitemShut {NoStop}%
\bibitem [{\citenamefont {{DeForest}}\ \emph {et~al.}(2013)\citenamefont
  {{DeForest}}, \citenamefont {{Howard}},\ and\ \citenamefont
  {{McComas}}}]{2013ApJ...769...43D}%
  \BibitemOpen
  \bibfield  {author} {\bibinfo {author} {\bibfnamefont {C.~E.}\ \bibnamefont
  {{DeForest}}}, \bibinfo {author} {\bibfnamefont {T.~A.}\ \bibnamefont
  {{Howard}}}, \ and\ \bibinfo {author} {\bibfnamefont {D.~J.}\ \bibnamefont
  {{McComas}}},\ }\href {\doibase 10.1088/0004-637X/769/1/43} {\bibfield
  {journal} {\bibinfo  {journal} {\apj}\ }\textbf {\bibinfo {volume} {769}},\
  \bibinfo {eid} {43} (\bibinfo {year} {2013})}\BibitemShut {NoStop}%
\bibitem [{\citenamefont {Amenomori}\ \emph {et~al.}(2013)\citenamefont
  {Amenomori} \emph {et~al.}}]{Amenomori:2013own}%
  \BibitemOpen
  \bibfield  {author} {\bibinfo {author} {\bibfnamefont {M.}~\bibnamefont
  {Amenomori}} \emph {et~al.} (\bibinfo {collaboration} {Tibet ASgamma}),\
  }\href {\doibase 10.1103/PhysRevLett.111.011101} {\bibfield  {journal}
  {\bibinfo  {journal} {Phys. Rev. Lett.}\ }\textbf {\bibinfo {volume} {111}},\
  \bibinfo {pages} {011101} (\bibinfo {year} {2013})},\ \Eprint
  {http://arxiv.org/abs/1306.3009} {arXiv:1306.3009 [astro-ph.SR]} \BibitemShut
  {NoStop}%
\bibitem [{\citenamefont {{Illing}}\ and\ \citenamefont
  {{Hundhausen}}(1985)}]{1985JGR....90..275I}%
  \BibitemOpen
  \bibfield  {author} {\bibinfo {author} {\bibfnamefont {R.~M.~E.}\
  \bibnamefont {{Illing}}}\ and\ \bibinfo {author} {\bibfnamefont {A.~J.}\
  \bibnamefont {{Hundhausen}}},\ }\href {\doibase 10.1029/JA090iA01p00275}
  {\bibfield  {journal} {\bibinfo  {journal} {\jgr}\ }\textbf {\bibinfo
  {volume} {90}},\ \bibinfo {pages} {275} (\bibinfo {year} {1985})}\BibitemShut
  {NoStop}%
\bibitem [{\citenamefont {{Panasenco}}\ \emph {et~al.}(2011)\citenamefont
  {{Panasenco}}, \citenamefont {{Martin}}, \citenamefont {{Joshi}},\ and\
  \citenamefont {{Srivastava}}}]{2011JASTP..73.1129P}%
  \BibitemOpen
  \bibfield  {author} {\bibinfo {author} {\bibfnamefont {O.}~\bibnamefont
  {{Panasenco}}}, \bibinfo {author} {\bibfnamefont {S.}~\bibnamefont
  {{Martin}}}, \bibinfo {author} {\bibfnamefont {A.~D.}\ \bibnamefont
  {{Joshi}}}, \ and\ \bibinfo {author} {\bibfnamefont {N.}~\bibnamefont
  {{Srivastava}}},\ }\href {\doibase 10.1016/j.jastp.2010.09.010} {\bibfield
  {journal} {\bibinfo  {journal} {Journal of Atmospheric and Solar-Terrestrial
  Physics}\ }\textbf {\bibinfo {volume} {73}},\ \bibinfo {pages} {1129}
  (\bibinfo {year} {2011})}\BibitemShut {NoStop}%
\bibitem [{\citenamefont {{Gopalswamy}}\ \emph {et~al.}(2003)\citenamefont
  {{Gopalswamy}}, \citenamefont {{Shimojo}}, \citenamefont {{Lu}},
  \citenamefont {{Yashiro}}, \citenamefont {{Shibasaki}},\ and\ \citenamefont
  {{Howard}}}]{2003ApJ...586..562G}%
  \BibitemOpen
  \bibfield  {author} {\bibinfo {author} {\bibfnamefont {N.}~\bibnamefont
  {{Gopalswamy}}}, \bibinfo {author} {\bibfnamefont {M.}~\bibnamefont
  {{Shimojo}}}, \bibinfo {author} {\bibfnamefont {W.}~\bibnamefont {{Lu}}},
  \bibinfo {author} {\bibfnamefont {S.}~\bibnamefont {{Yashiro}}}, \bibinfo
  {author} {\bibfnamefont {K.}~\bibnamefont {{Shibasaki}}}, \ and\ \bibinfo
  {author} {\bibfnamefont {R.~A.}\ \bibnamefont {{Howard}}},\ }\href {\doibase
  10.1086/367614} {\bibfield  {journal} {\bibinfo  {journal} {\apj}\ }\textbf
  {\bibinfo {volume} {586}},\ \bibinfo {pages} {562} (\bibinfo {year}
  {2003})}\BibitemShut {NoStop}%
\bibitem [{\citenamefont {{Yashiro}}\ \emph {et~al.}(2004)\citenamefont
  {{Yashiro}}, \citenamefont {{Gopalswamy}}, \citenamefont {{Michalek}},
  \citenamefont {{St.~Cyr}}, \citenamefont {{Plunkett}}, \citenamefont
  {{Rich}},\ and\ \citenamefont {{Howard}}}]{2004JGRA..109.7105Y}%
  \BibitemOpen
  \bibfield  {author} {\bibinfo {author} {\bibfnamefont {S.}~\bibnamefont
  {{Yashiro}}}, \bibinfo {author} {\bibfnamefont {N.}~\bibnamefont
  {{Gopalswamy}}}, \bibinfo {author} {\bibfnamefont {G.}~\bibnamefont
  {{Michalek}}}, \bibinfo {author} {\bibfnamefont {O.~C.}\ \bibnamefont
  {{St.~Cyr}}}, \bibinfo {author} {\bibfnamefont {S.~P.}\ \bibnamefont
  {{Plunkett}}}, \bibinfo {author} {\bibfnamefont {N.~B.}\ \bibnamefont
  {{Rich}}}, \ and\ \bibinfo {author} {\bibfnamefont {R.~A.}\ \bibnamefont
  {{Howard}}},\ }\href {\doibase 10.1029/2003JA010282} {\bibfield  {journal}
  {\bibinfo  {journal} {Journal of Geophysical Research (Space Physics)}\
  }\textbf {\bibinfo {volume} {109}},\ \bibinfo {eid} {A07105} (\bibinfo {year}
  {2004})}\BibitemShut {NoStop}%
\bibitem [{\citenamefont {{Low}}(1996)}]{1996SoPh..167..217L}%
  \BibitemOpen
  \bibfield  {author} {\bibinfo {author} {\bibfnamefont {B.~C.}\ \bibnamefont
  {{Low}}},\ }\href {\doibase 10.1007/BF00146338} {\bibfield  {journal}
  {\bibinfo  {journal} {\solphys}\ }\textbf {\bibinfo {volume} {167}},\
  \bibinfo {pages} {217} (\bibinfo {year} {1996})}\BibitemShut {NoStop}%
\bibitem [{\citenamefont {{Endeve}}\ \emph {et~al.}(2004)\citenamefont
  {{Endeve}}, \citenamefont {{Holzer}},\ and\ \citenamefont
  {{Leer}}}]{2004ApJ...603..307E}%
  \BibitemOpen
  \bibfield  {author} {\bibinfo {author} {\bibfnamefont {E.}~\bibnamefont
  {{Endeve}}}, \bibinfo {author} {\bibfnamefont {T.~E.}\ \bibnamefont
  {{Holzer}}}, \ and\ \bibinfo {author} {\bibfnamefont {E.}~\bibnamefont
  {{Leer}}},\ }\href {\doibase 10.1086/381239} {\bibfield  {journal} {\bibinfo
  {journal} {\apj}\ }\textbf {\bibinfo {volume} {603}},\ \bibinfo {pages} {307}
  (\bibinfo {year} {2004})}\BibitemShut {NoStop}%
\bibitem [{\citenamefont {{Fujiki}}\ \emph {et~al.}(2016)\citenamefont
  {{Fujiki}}, \citenamefont {{Tokumaru}}, \citenamefont {{Hayashi}},
  \citenamefont {{Satonaka}},\ and\ \citenamefont
  {{Hakamada}}}]{2016ApJ...827L..41F}%
  \BibitemOpen
  \bibfield  {author} {\bibinfo {author} {\bibfnamefont {K.}~\bibnamefont
  {{Fujiki}}}, \bibinfo {author} {\bibfnamefont {M.}~\bibnamefont
  {{Tokumaru}}}, \bibinfo {author} {\bibfnamefont {K.}~\bibnamefont
  {{Hayashi}}}, \bibinfo {author} {\bibfnamefont {D.}~\bibnamefont
  {{Satonaka}}}, \ and\ \bibinfo {author} {\bibfnamefont {K.}~\bibnamefont
  {{Hakamada}}},\ }\href {\doibase 10.3847/2041-8205/827/2/L41} {\bibfield
  {journal} {\bibinfo  {journal} {\apjl}\ }\textbf {\bibinfo {volume} {827}},\
  \bibinfo {eid} {L41} (\bibinfo {year} {2016})}\BibitemShut {NoStop}%
\bibitem [{\citenamefont {{Hathaway}}\ and\ \citenamefont
  {{Upton}}(2016)}]{2016JGRA..12110744H}%
  \BibitemOpen
  \bibfield  {author} {\bibinfo {author} {\bibfnamefont {D.~H.}\ \bibnamefont
  {{Hathaway}}}\ and\ \bibinfo {author} {\bibfnamefont {L.~A.}\ \bibnamefont
  {{Upton}}},\ }\href {\doibase 10.1002/2016JA023190} {\bibfield  {journal}
  {\bibinfo  {journal} {Journal of Geophysical Research (Space Physics)}\
  }\textbf {\bibinfo {volume} {121}},\ \bibinfo {pages} {10} (\bibinfo {year}
  {2016})},\ \Eprint {http://arxiv.org/abs/1611.05106} {arXiv:1611.05106
  [astro-ph.SR]} \BibitemShut {NoStop}%
\bibitem [{\citenamefont {Agostinelli}\ \emph {et~al.}(2003)\citenamefont
  {Agostinelli} \emph {et~al.}}]{Agostinelli:2002hh}%
  \BibitemOpen
  \bibfield  {author} {\bibinfo {author} {\bibfnamefont {S.}~\bibnamefont
  {Agostinelli}} \emph {et~al.} (\bibinfo {collaboration} {GEANT4}),\ }\href
  {\doibase 10.1016/S0168-9002(03)01368-8} {\bibfield  {journal} {\bibinfo
  {journal} {Nucl. Instrum. Meth.}\ }\textbf {\bibinfo {volume} {A506}},\
  \bibinfo {pages} {250} (\bibinfo {year} {2003})}\BibitemShut {NoStop}%
\bibitem [{\citenamefont {{Burlaga}}\ \emph {et~al.}(2002)\citenamefont
  {{Burlaga}}, \citenamefont {{Ness}}, \citenamefont {{Wang}},\ and\
  \citenamefont {{Sheeley}}}]{2002JGRA..107.1410B}%
  \BibitemOpen
  \bibfield  {author} {\bibinfo {author} {\bibfnamefont {L.~F.}\ \bibnamefont
  {{Burlaga}}}, \bibinfo {author} {\bibfnamefont {N.~F.}\ \bibnamefont
  {{Ness}}}, \bibinfo {author} {\bibfnamefont {Y.-M.}\ \bibnamefont {{Wang}}},
  \ and\ \bibinfo {author} {\bibfnamefont {N.~R.}\ \bibnamefont {{Sheeley}}},\
  }\href {\doibase 10.1029/2001JA009217} {\bibfield  {journal} {\bibinfo
  {journal} {Journal of Geophysical Research (Space Physics)}\ }\textbf
  {\bibinfo {volume} {107}},\ \bibinfo {eid} {1410} (\bibinfo {year}
  {2002})}\BibitemShut {NoStop}%
\bibitem [{\citenamefont {{Strauss}}\ \emph {et~al.}(2012)\citenamefont
  {{Strauss}}, \citenamefont {{Potgieter}}, \citenamefont {{B{\"u}sching}},\
  and\ \citenamefont {{Kopp}}}]{2012Ap&SS.339..223S}%
  \BibitemOpen
  \bibfield  {author} {\bibinfo {author} {\bibfnamefont {R.~D.}\ \bibnamefont
  {{Strauss}}}, \bibinfo {author} {\bibfnamefont {M.~S.}\ \bibnamefont
  {{Potgieter}}}, \bibinfo {author} {\bibfnamefont {I.}~\bibnamefont
  {{B{\"u}sching}}}, \ and\ \bibinfo {author} {\bibfnamefont {A.}~\bibnamefont
  {{Kopp}}},\ }\href {\doibase 10.1007/s10509-012-1003-z} {\bibfield  {journal}
  {\bibinfo  {journal} {\apss}\ }\textbf {\bibinfo {volume} {339}},\ \bibinfo
  {pages} {223} (\bibinfo {year} {2012})}\BibitemShut {NoStop}%
\bibitem [{\citenamefont {Cholis}\ \emph {et~al.}(2016)\citenamefont {Cholis},
  \citenamefont {Hooper},\ and\ \citenamefont {Linden}}]{Cholis:2015gna}%
  \BibitemOpen
  \bibfield  {author} {\bibinfo {author} {\bibfnamefont {I.}~\bibnamefont
  {Cholis}}, \bibinfo {author} {\bibfnamefont {D.}~\bibnamefont {Hooper}}, \
  and\ \bibinfo {author} {\bibfnamefont {T.}~\bibnamefont {Linden}},\ }\href
  {\doibase 10.1103/PhysRevD.93.043016} {\bibfield  {journal} {\bibinfo
  {journal} {Phys. Rev.}\ }\textbf {\bibinfo {volume} {D93}},\ \bibinfo {pages}
  {043016} (\bibinfo {year} {2016})},\ \Eprint
  {http://arxiv.org/abs/1511.01507} {arXiv:1511.01507 [astro-ph.SR]}
  \BibitemShut {NoStop}%
\bibitem [{\citenamefont {{Gopalswamy}}\ \emph {et~al.}(2016)\citenamefont
  {{Gopalswamy}}, \citenamefont {{Yashiro}},\ and\ \citenamefont
  {{Akiyama}}}]{2016ApJ...823L..15G}%
  \BibitemOpen
  \bibfield  {author} {\bibinfo {author} {\bibfnamefont {N.}~\bibnamefont
  {{Gopalswamy}}}, \bibinfo {author} {\bibfnamefont {S.}~\bibnamefont
  {{Yashiro}}}, \ and\ \bibinfo {author} {\bibfnamefont {S.}~\bibnamefont
  {{Akiyama}}},\ }\href {\doibase 10.3847/2041-8205/823/1/L15} {\bibfield
  {journal} {\bibinfo  {journal} {\apjl}\ }\textbf {\bibinfo {volume} {823}},\
  \bibinfo {eid} {L15} (\bibinfo {year} {2016})},\ \Eprint
  {http://arxiv.org/abs/1605.02217} {arXiv:1605.02217 [astro-ph.SR]}
  \BibitemShut {NoStop}%
\end{thebibliography}%

\end{document}